\documentclass{llncs}
\usepackage{type1cm}        
\usepackage{array, makecell}
\usepackage[pdftex]{graphicx}       
\graphicspath{{./figs/}}    
\usepackage{subfig}         

\usepackage{multirow}
\usepackage{amsfonts}
\usepackage[cmex10]{amsmath}
\usepackage[utf8]{inputenc}
\usepackage[english]{babel}
\usepackage{xcolor}
\usepackage[square,numbers]{natbib}

\usepackage{rotating}

\usepackage{booktabs} 
\usepackage{soul}
\usepackage[flushleft]{threeparttable}
\usepackage{changepage} 
\usepackage{lscape}
\usepackage[hyphens]{url}
\usepackage[hyphenbreaks]{breakurl}
\makeatletter 
\renewcommand\@biblabel[1]{#1.} 
\makeatother

\usepackage[final]{changes}

\begin{document}
\title{Group Threat, Political Extremity, and Collective Dynamics in Online Discussions}
\titlerunning{Collective Dynamics in Online Discussions} 
%
\author{N. Gizem Bacaksizlar Turbic*\inst{1,2,3} \and Mirta Galesic \inst{1}}
\authorrunning{Bacaksizlar Turbic and Galesic} 
\institute{\textsuperscript{1}Santa Fe Institute, Santa Fe, New Mexico, 87501, United States\\
\textsuperscript{2}NOVA University of Lisbon, Lisbon, 1069-061, Portugal\\
\textsuperscript{3}GESIS - Leibniz Institute for the Social Sciences, Cologne, 50667, Germany\\
\email{{gizem.bacaksizlarturbic@gesis.org}, {galesic@santafe.edu}}}
%

\maketitle    
\section*{Abstract}
Collectives adapt their network structure to the challenges they face. It has been hypothesized that collectives experiencing a real or imagined threat from an outgroup tend to consolidate behind a few \deleted{influential} group members, and that network structures in which a few members \added{attract most of the attention} \deleted{have a very strong influence} are more likely in politically extreme groups. These hypotheses have not been tested in large-scale real-world settings. We reconstruct networks of tens of thousands of commenters participating in comment sections of high-profile U.S. political news websites spanning the political spectrum from left to right, including Mother Jones, The Atlantic, The Hill, and Breitbart. We investigate the relationship between different indices of inequality of \deleted{influence} \added{attention} in commenters' networks and perceived group threat associated with significant societal events, from elections and political rallies to mass shootings. Our findings support the hypotheses that groups facing a real or imagined outgroup threat and groups that are more politically extreme are more likely to \deleted{include disproportionately influential} \added{attend to a few high-profile members}. These results provide an extensive real-world test of theoretical accounts of collective adaptation to outgroup threats. 

\section*{Introduction}
New technologies \deleted[id=r2]{enable quick}\added[id=r2]{facilitate} formation and dissolution of network connections, allowing collectives to quickly consolidate behind ideas and leaders they find useful for solving their perceived \added[id=r2]{and real} problems \cite{lorenz2021digital}. While this collective dynamics \added[id=r2]{is not new}\deleted[id=r2]{was always at play}, online communication platforms make it particularly easy for some \added[id=r2]{individuals}\deleted[id=r2]{ members} to \added[id=r2]{attract collective attention}. \added[id=r2]{While not sufficient, the attention is a necessary} \deleted{the pre}condition for any later influence one might have on the discourse, beliefs, and behaviors of others in the collective. More prominent members have the power to steer the conversation \cite{lewandowsky2020using} and influence others through both normative conformity, by popularizing specific worldviews and behaviors, and informational conformity, \deleted[id=r2]{where their followers trust them}\added[id=r2]{by being trusted} as sources of reliable information \cite{claidiere2012integrating}. Given the frequent disconnect between the quality of arguments and their popularity in online media, \cite{bak2021stewardship}, such dynamics can have consequences that are ultimately harmful to collectives and their members \cite{jennings2021lack}. 

\added{Here we investigate collective dynamics in comment sections of news sites, where } millions of people comment daily on current societal events\deleted[id=r2]{using a variety of online platforms, from social media to comment sections of news websites} \cite{ziegele2018socially}. Reading others' comments can shape one's own opinions about an event and the source of the story \cite{prochazka2018effects}, and help spread opinions and claims which counter the mainstream narrative \cite{toepfl2015public}. Commenters can influence how other readers perceive the issue being discussed in the article, their evaluations of the article's quality, and their affective response to the issue \cite{ziegele2014creates,naab2020comments}. While participants in public discourse should ideally have similar opportunities to be heard \cite{habermas1991structural, hauser2022vernacular}, in online discussions \added[id=r2]{some commenters attract more attention than others and are more likely to influence the subsequent discussion}. 

\added[id=r2]{We reconstruct networks of commenters discussing major societal events, from elections to political rallies and mass shootings, on four prominent news websites spanning the U.S. political spectrum from left to right: Mother Jones, The Atlantic, The Hill, and Breitbart. We study conditions under which these commenter networks consolidate around a smaller set of high-profile voices, as reflected in the unequal distribution of replies commenters get from others \cite{ziegele2018dynamics}.} 

Two socio-psychological factors, in particular, have been proposed to promote unequal \deleted{influence of} \added{attention to} group members: imagined or real outgroup threat, and political extremity of the group. \deleted[id=r2]{Here we study how these factors relate to the inequality of \deleted{influence} \added{attention} in large collectives of commenters.  \deleted{The inequality of influence is a multifaceted concept that can be investigated in different ways. For example, an influential commenter can be described as someone who can steer the conversation, influence others' beliefs and behaviors, or simply as someone attracting more attention.} \deleted{Here}We focus on attention\deleted{, as} signaled by replies commenters get from others \cite{ziegele2018dynamics}\deleted{, because}. Based on reply patterns, we reconstruct networks of commenters discussing major societal events, from elections to political rallies and mass shootings, on four prominent news websites spanning the U.S. political spectrum from left to right: Mother Jones, The Atlantic, The Hill, and Breitbart.}We measure how different individual- and network-based indices of inequality of \deleted{influence}\added{attention} change around societal events perceived as more or less threatening and how they differ for more or less politically extreme commenter collectives.

\textbf{Group threat and inequality of \deleted{influence}\added{attention}.} Real or imagined \emph{group threat} from an opposing other group (outgroup) to one's own group (ingroup) can prompt a series of changes in group dynamics that might help a group to consolidate and coordinate for better defense or attack. Specifically, a threat from another group can lead to a decrease in diversity of opinions within a group and a higher likelihood of following opinions of a few influential group members \cite{turner1992threat}. This is particularly likely for groups under threat that are also highly cohesive \cite{festinger1950social, thompson1976effects}, with members that share similar characteristics and views \cite{colleoni2014echo}, and are motivated to protect the collective positive image \cite{raven1974nixon}. This reaction can be adaptive because groups that are more cohesive and coordinated can have an advantage in inter-group conflicts \cite{kirke2010military,maccoununit}, but it can also lower groups' ability to solve problems that require diversity of opinions \cite{janis1982groupthink}. 

We, therefore, hypothesize that inequality of commenters' \deleted{influence}\added{attention} will become larger after events that \deleted{lead to an increase of}\added{bring} a real or imagined threat to one's ingroup from a specific outgroup. For example, just before the 2016 U.S. Presidential Election, Clinton was favored as the election winner, and this might have been perceived as a threat among the Trump-supporting voters. After the 2016 Election and Trump's unexpected victory, the Clinton-supporters might have felt threatened. Consequently, we predict higher inequality of \deleted{influence}\added{attention} among commenters on right-wing sites (such as Breitbart) before the 2016 U.S. Election, and among commenters on the left-oriented ones (such as Mother Jones and The Atlantic) after the Election. A similar change could be expected after the 2017 Trump Presidential Inauguration, which consolidated his victory, and after the alt-right 2017 Charlottesville Rally. The same type of change, but in the opposite direction, could have happened after the 2018 U.S. Election for Congress, where Democrats achieved significant victories. 

\textbf{Political extremity and inequality of \deleted{influence}\added{attention}.} Another factor \added{that could be} related to inequality of \deleted{influence in a group}\added{attention to other group members} is \emph{political extremity} of a collective. Both radical right- and radical left-leaning political extremists have been linked to heightened deference to authorities compared to more moderate political ideologies \cite{altemeyer1996authoritarian,shils1954authoritarianism}. The propensity towards authoritarianism has traditionally been linked to right-wing ideology and operationalized in terms of ideological commitment to tradition, authority, and social convention against threats of change, protest, and political rebellion \cite{adorno1950theauthoritarian,altemeyer1998other, jost2003political}. However, members of left-oriented groups, especially those who label themselves as communist, can also show authoritarian traits \cite{de2011left}. While overtly the extreme right- and left-oriented groups differ in significant ways, such as in their attitudes towards the existing social order vs. liberation of oppressed groups, they also show important similarities, including estrangement from the government, intolerance of ambiguity, intolerance towards political opponents, attraction to totalitarian measures and tactics, intolerance of human frailty, and paranoid tendencies - including a belief in conspiracy and feelings of persecution  \cite{mcclosky1985similarities,stone1980myth}. 
Consequently, we hypothesize that networks of commenters on news sites advocating more radical political beliefs (both left-oriented ones, such as Mother Jones and The Atlantic, and the right-oriented ones, such as Breitbart; \cite{adfontes, allsides,biasly, mediabiascheck}), which likely include commenters with similar levels and directions of political extremity, will be characterized by a larger inequality of \deleted{influence}\added{attention} than the commenter networks on the more centrist news sites (such as The Hill).

\textbf{Inequality of \deleted{influence}\added{attention} in real-world discussions.} To investigate how well these theoretical predictions describe the actual public discourse, we collected two different data sets. The first one consists of all comments posted on Mother Jones, The Atlantic, The Hill, and Breitbart news websites a month before and a month after seven important societal and political events, and the information about who replies to whom (see Method for details). We include all political events ranked among the top five U.S. news according to Google Trends from 2016 to 2018 (Table \ref{tab:events}, \cite{trends}). In addition, we include the 2017 Presidential Inauguration and the 2017 Charlottesville Rally, which were not listed among the top five news but were highly covered due to their wide-reaching political and social effects \cite{history,npr}. Of the seven events, four involved a clear conflict between different political groups and could have been perceived as a real or imagined threat from a specific other group: the 2016 U.S. Election, the 2017 Presidential Inauguration, the 2017 Charlottesville alt-right Rally, and the 2018 U.S. Election. The other three events were also societally very important but the group that caused the threat was less specific, at least for the commenters on the U.S. left- and right-oriented news websites: the 2016 Orlando terrorist attack, the 2016 Brexit Referendum, and the 2017 Las Vegas \added[id=r2]{Shooting}. 

We recreate the network of commenters before and after each event, where links are direct replies from one commenter to another, across different discussions, weighted by the overall number of replies in a particular time period. For each news website, for the month before and after each societal event, we \added{derived the network of commenters and} calculated four measures of \deleted{influence}\added{attention} inequality in \deleted{networks}\added{these} networks. Two of the measures describe inequality of \deleted{influence}\added{attention} of individual commenters, namely, 1) skewness of commenters' weighted in-degree distribution and 2) skewness of the distribution of their Page Rank centralities. Presumably, people who reply to a commenter have \deleted{read}\added{attended to} their comment, which then can have some influence on their own thinking, even if they disagree with the comment. The more unequal this distribution of in-degrees, the more unequal the \added{attention to and the} potential influence of different commenters. Similarly, a commenter that is replied to by commenters who themselves have a lot of replies (i.e., who has a higher Page Rank) could be considered to have a higher \added{potential} influence. The other two measures describe the aspects of the overall structure of commenter networks that are related to inequality \added{of attention}, namely, 3) the proportion of independent commenters who neither reply to others nor are replied to, and 4) the proportion of connected components of the network. The proportion of independent commenters can be expected to decrease under real or imagined threat as commenters \deleted{might more often respond to the influential}\added{consolidate behind a few prominent} users, rather than posting alone. Similarly, if people preferentially reply to a smaller subset of commenters\deleted{, who can therefore be viewed as more influential,} there should be relatively fewer connected components in the network than if there is no particular preference for replying to some commenters.

Our second data set includes human judgments about the perceived threat posed by different events, which we collected in a survey of participants with different political views (see Method for details). This data allowed us to estimate how much   \deleted{were} the chosen events \added{were} perceived as threatening to one's own ingroup, depending on the\deleted{ir} \added{ingroup's} political orientation.

We use these data sets to investigate our two main questions: 1) How does inequality of \deleted{influence}\added{attention} change when a group is experiencing a threat from an outgroup?, and 2) How does the inequality of \deleted{commenters' influence}\added{attention} differ for more or less politically extreme news sites?

\section*{Method}

\textbf{Data on commenter networks around important societal events.} We collected all comments posted a month before and a month after seven important societal events, through Disqus API (the data collection took place in 2019). The month-long interval before and after each event provided us with sufficient data from all sites, while not complicating the analysis with other significant events occurring in a temporal vicinity. Table \ref{tab:counts} presents the number of unique commenters, comments, and articles included in the analysis. Figure \ref{fig:comart_distributions} in Supplementary Material shows the number of comments per article for different websites and events. 

\textbf{Events.} Four of the seven events involve a clearly defined ingroup and outgroup for most people in the U.S. and are therefore expected to evoke more perceived outgroup threat. As described in the Introduction, such events are the 2016 U.S. Presidential Election, the 2017 Trump Presidential Inauguration, the alt-right 2017 Charlottesville Rally, and the 2018 U.S. Election for Congress. The other three events are less clearly connected to a specific outgroup, at least in the U.S. context, and are expected to provoke a lower perceived outgroup threat. The other two are perpetrated by single individuals without clear motivation. The 2016 Orlando Shooting was committed by a disturbed individual that might have attacked the gay Hispanic community but could have also been driven by Islamic terrorism. In the 2017 Las Vegas Shooting, a single person attacked victims without any discrimination. The last event is a shock happening to another society: the 2016 Brexit, where most Americans did not clearly belong to either of the two main groups involved. After these events, the whole society might feel threatened but not \added{be} able to pinpoint the threat to a clearly defined outgroup. Instead, this generalized feeling of threat might prompt more information-seeking and questioning of existing views, division into smaller groups to explore a broader range of possible solutions, and seeking information outside the group \cite{marcus1993anxiety,brader2005striking,valentino2008worried,redlawsk2010affective}; though see \cite{keinan1987decision,sengupta2001contingent}. Therefore, such unspecific threats may lead to a lower rise in inequality within a group than the more specific outgroup threats. 

\textbf{Websites.} We collected comments from discussion sections of four U.S. news websites, including two on the left side of the political spectrum - Mother Jones and The Atlantic, one centrist or moderate right - The Hill, and one on the extreme right - Breitbart. We derived the political ideology of these news websites according to several different media rating systems at the time of the data collection \cite{adfontes, allsides,biasly, mediabiascheck}. Mother Jones \cite{motherjones} is a magazine launched in 1976 that includes news, commentary, and investigative reporting on topics ranging from politics and the environment to health and culture. According to the rating systems,  Mother Jones typically espouses liberal or progressive views. The Atlantic \cite{atlantic} is a magazine founded in 1857, providing daily coverage and analysis of news and issues related to politics, education, technology, health, science, and culture. Its political inclination is typically rated as left or left-center. The Hill \cite{thehill} is a news website established in 1994, focusing on politics, policy, business, and international relations. During the time period, we focus on here, it has been generally rated as the right of center \cite{adfontes,biasly}. Breitbart \cite{breitbart}, founded in 2007, is a news, opinion, and commentary website. It is generally considered to have an extreme right bias and to be an unreliable news source (e.g., \cite{wiki_breitbart}, though see \cite{fb_breitbart}) as it sometimes publishes articles in support of conspiracy theories and false claims.

For the purpose of this analysis, we assume that most commenters on a particular news website have political leanings aligned with those advocated by the website. While this is certainly not the case for all commenters, our extensive reading of the comments posted on each of these websites suggests that this holds in a vast majority of cases. \added{In addition, we have analyzed the frequency with which commenters who posted mostly on one site also commented on the other news sites. As shown in Tables \ref{tab:overlaps1} and \ref{tab:overlaps2}, the percentage of such commenters and their comments is quite small.}

During most of the period studied in this paper (2016-2018), all four news websites used the Disqus platform to allow readers to comment on their  articles \cite{disqus}. The Atlantic removed commenting sections in early 2018, and Mother Jones switched to another commenting platform by mid-2019, ending the possibility of comparing Disqus comments on left vs. right news sites. The Hill and Breitbart used Disqus as their commenting platform throughout the studied period.

The comments from the news websites were obtained within the terms of use of the Disqus API (see \cite{disquspolicy}, in particular, the section ``Rights Regarding User Content"), and Disqus displayed its privacy policy prominently at the top of the commenting section of each site \cite{disqushelp}. This research activity is exempt from requiring IRB approval because the comments are publicly available and used in a completely anonymized form (see paragraph 46.104-d-4 at \cite{IRB}).

\textbf{Establishing commenter networks.} For each comment, we have information about the unique identification number (id) of the commenter who posted it and whether the comment was a reply to another comment (i.e., whether the comment was a `child' of a `parent' comment) or posted to the general discussion (and then either attracted others' replies and, becoming a `parent' of a thread of comments, or not getting any replies, thus remaining fully `independent'). To post comments, readers have to register for Disqus but can otherwise remain anonymous. Disqus does not require linking one's account to a social media account or another form of identification, and \deleted{our extensive reading of the comment sections suggests that the vast majority of}\added{indeed we observe that most} Disqus users commenting on the websites we study do not use their real names. Nevertheless, because commenters need to register, each comment can be uniquely attributed to a specific individual. 

Using this information, we can build a commenter network. Within a specific period of time, commenters can be represented as a part of a \textit{directed network} in which edges are direct replies from one commenter to the other, accumulated over that period across different articles and comment threads. The weight of a directed link is equal to the number of replies from one commenter to the other. Independent commenters, who neither reply to anyone nor are replied to by others, are represented as disconnected nodes in this network. 
 
Besides posting comments, two other ways of interacting with comment sections are possible. Registered readers can simply upvote or downvote a comment. Unlike comments, these votes cannot be traced back to specific individuals and are therefore not useful for the purpose of building a network of commenters. In addition, some news websites allow readers to react to certain articles through the websites' social media accounts. These comments - typically 10-15\% of all comments - appear in comment sections as `guest' comments. The individual authors of these comments cannot be traced, and therefore they are not included in our analyses. 
 
Some commenters get more replies than others (Figure \ref{fig:indegree_distributions}), for at least two intertwined reasons. First, certain commenters are very popular \added{to begin with} and other readers might be particularly alert and willing to reply to their comments. Second, some commenters' comments can be more visible due \added{to} the way comments are \deleted{sorted}\added{displayed}. \added{On social media sites such as Facebook and Twitter, numerous algorithms control the display, making it more likely that users see other users, messages, and promotions that are in line with their personal preferences. In the comment sections of the news sites that we investigate, there is no such customization. Users can change the display in only two ways: they can sort the comments in different ways, and they can choose how many comments at the time to display.} A default sorting in Disqus is a function of \deleted{the} the number of up- and downvotes received as the parent comment in a thread, or as an independent comment (`best' sorting \cite{disqussort}). Users can instead choose to sort the comments from the `newest' to the `oldest', or vice versa. They can see the first 50 comments according to the current sorting and typically have to press the `Load more comments' button at the bottom of the page if they wish to see further comments. Obviously, the `best' sorting will lead to a `rich get richer' dynamics, whereby initially more upvoted comments will be more visible, potentially attracting additional votes and replies. These two factors are interrelated: more popular commenters are likely to get more upvotes early on, which promotes their visibility and further promotes their popularity. 

\added{In addition, site moderators can affect somewhat what the users can see. They can set up more or less strict moderation settings and change them over time \cite{disqusmod}. For example, they can automatically delete comments containing certain words or allow commenting only by registered users. They can also allow users to upvote and/or downvote comments. These uncontrollable differences are likely to introduce noise in our analyses but are unlikely to cause systematic biases. Overall, the fact that all four news sites we investigate use the same commenting platform enables an easier comparison of commenting patterns across sites because the commenting interface is the same on all of them.} 

We can use the commenter network to investigate different aspects of inequality of \deleted{influence}\added{attention} among commenters. Here we focus on network-based indices, including those measuring the skewness of the distribution \deleted{of individual influence}\added{at the level of individual commenters} (the first two measures below), and the inequality at the level of the overall network structure (the second two measures below): 

\begin{enumerate}
\item \textit{Skewness of the weighted in-degree distributions.} 
For each commenter, we calculate their weighted in-degree - the sum of the link weights (i.e., number of replies from other commenters) for links incident to that commenter in a given period. We normalize in-degrees within each period to the range [0 1], where 0 corresponds to the lowest value in the data set and 1 to the highest one. To get an inequality of the in-degree distribution \cite{newman2004analysis}, we calculate \added{the} skewness index, corrected for network size \added{(equations 1 and 5 in \cite{bobee1975correction}, implemented in Matlab package `skewness')}. Note that skewness is more appropriate than the Gini index \cite{bendel1989comparison} when the most important changes are at the right-hand tail of the distribution, as is the case here. As we are particularly interested in between-group differences and temporal changes at the tail ends of these distributions, we calculate skewness for the top 1\% of commenters ordered by centrality, as well as for all commenters together. We expect clearer results for the top 1\% because distributions of centralities have very long tails (see Figures \ref{fig:indegree_distributions} and \ref{fig:pagerank_distributions} in Supplementary Material) that can dampen measures of change in the inequality occurring at the very top part of the distributions.

\item \textit{Skewness of Page Rank distributions.} 
For each commenter, we calculate their Page Rank index, normalize all indices within each period to [0 1] range as above, and assess the inequality of Page Rank \cite{page1999pagerank} distributions in the same way as described above for in-degree distributions.

\item \textit{Proportion of the independent commenters among all commenters.} Independent commenters are those that neither reply to another commenter (they always post their comments directly to the general discussion) nor are replied to by other commenters. 
 
\item \textit{Proportion of connected components in the commenter network.} We calculated the number of connected components and divided it by the maximum number of components, that is, the total number of commenters.

\end{enumerate}

Another way of measuring \deleted{influence}\added{attention} could be the average size of a reply tree that a commenter managed to initiate in a particular time period. However, to the readers it is often not clear who initiated a specific discussion, because of the limited number of comments (50) that can be shown on a single screen, and \added{because of} the condensed way in which various branches of the reply tree are shown\deleted{, at least on the news websites we study}. Most discussions occur between commenters within a particular part of the tree, rather than between the top commenter and the other commenters. Therefore, in our analysis, we focus only on the direct links between commenters, rather than counting all indirect links. 

\textbf{Data on perceived threats associated with important societal events.} In addition to the network data, we collected evaluations of different aspects of perceived threat evoked by each of the seven societal events described above, in a survey on a sample of 100 participants from Amazon Mechanical Turk \cite{mturk}. The median age of participants was 38 (24-78), with 55 of them identifying as men and 45 as women, and 73 having college degrees. We used the service CloudResearch \cite{chandler2019online} to select participants who previously identified themselves as having liberal or conservative political views. We asked them about their political orientation and found that 55 reported liberal political views, 40 conservative, and 5 moderate views. These sample sizes were determined as sufficient to reliably detect the expected large differences between liberal- and conservative-leaning participants in the evaluation of threats from different political events \cite{dimock2014polarization,doherty2016security,tyson2020threat}. All participants provided an informed consent and the study was approved by the IRB of the University of New Mexico (1331148). All methods were performed in accordance with the relevant guidelines and regulations.

Participants answered seven questions about each event. Two questions asked about the threat they experienced personally: ``How threatened did you personally feel in the month after this event?" \textit{(1-Not at all threatened...7-Extremely threatened)}; and ``In the month after the event, did you feel more or less threatened compared to the month before the event?" \textit{(1-I felt much less threatened...7-I felt much more threatened)}. One question asked about the overall perceived threat for most people: ``Thinking now of all people in the U.S., independently of their political orientation: In the month after the event, did most people feel more or less threatened compared to the month before the event?" \textit{(1-Most people felt much less  threatened...7-Most people felt much more threatened)}.

Most importantly for the purpose of this study, four questions measured perceived threat to one's own ingroup and outgroup: ``In your view, did the event clearly benefit Democrats at the expense of Republicans, Republicans at the expense of Democrats, or were both groups affected similarly?" \textit{(The event benefited Democrats at the expense of Republicans; The event benefited Republicans at the expense of Democrats; The event benefited both Democrats and Republicans; The event hurt both Democrats and Republicans)}; ``Who do you think felt more threatened in the month after this event, Democrats or Republicans?" \textit{(1-Democrats felt much more threatened...7-Republicans felt much more threatened)}; ``Thinking now of Democrats, do you think they felt more or less threatened in the month after the event compared to the month before this event?" \textit{(1-Democrats felt much less threatened...7-Democrats felt much more threatened)}; and ``Thinking now of Republicans, do you think they felt more or less threatened in the month after the event compared to the month before this event?" \textit{(1-Republicans felt much less threatened...7-Republicans felt much more threatened)}. 

\added{As expected, for most specific threats, there is a larger difference between threat experienced by our liberal and conservative participants, compared to non-specific threats (panel 1 of Figure \ref{fig:threat_perception} in Supplementary Material). Similarly, for change in threat from before to after the event, the differences between our liberal and conservative participants are typically larger for specific threats than for non-specific threats (panel 2). For all specific threats, all participants (both liberal and conservative) say one party benefits at the expense of the other more often, whereas for all non-specific threats they more often say that both parties benefit or both parties hurt (panel 3). More participants say that one party feels more threatened by specific threats, compared to non-specific threats where more participants say that both parties feel threatened (panel 4). Participants evaluated that Democrats felt more threatened after compared to before the 2016 U.S. Election, the 2017 Inauguration, and the 2017 Charlottesville Rally; and less threatened after compared to before the 2018 U.S. Election (panel 5). Their evaluations are in the opposite direction for Republicans, as expected (panel 6). For non-specific threats, differences are smaller and the same for both parties (panels 5 and 6). There is a stronger perceived decrease in threat after Brexit for Republicans than for Democrats, but for both groups, the change is in the same direction, as expected. Finally, on average there is more difference between our liberal and conservative participants in perceived threat to all people in the U.S. for specific than for non-specific group threats.}

To control for the influence of distortions in memory since the event, for each event, participants also reported how well they remembered it \textit{(1-Hardly remember it at all...7-Remember it very vividly)}. The temporal distance was not predictive of the level of recall: the lowest rating (4.4 for liberal and 4.5 for conservative participants) was given to the 2018 U.S. Election, while the highest rating was received by the 2016 U.S. Election (5.9 for liberal and 5.8 for conservative participants). The level of recall did correlate with ratings of threat: on almost all questions, participants who reported recalling the event better also reported a higher level of threat (on average, liberal participants who recalled an event better perceived its threat across different questions about 0.36 scale points higher; and 0.32 for conservative participants). All our results remain consistent when analyzed separately for participants with better or worse reported recall, so for completeness, we report the analyses based on all participants. The complete results also likely reflect the full range of diversity of reactions to different events among commenters in the real world. 

\section*{Results}

\deleted{Overall, we find a high level of inequality among commenters n terms of various centrality measures, reflected in a high positive skew of the distributions of in-degree and Page Rank indices for different commenters (Figures~\ref{fig:indegree_distributions} and \ref{fig:pagerank_distributions} in Supplementary Material).}\added{Before describing the results, we should note that} all results have a high level of variability on several levels, across news websites, time periods, and different measures \added{of inequality}. However, some clear trends emerge in the answers to both of our research questions. 

\begin{figure}
    \centering
        \includegraphics[width=1\textwidth]{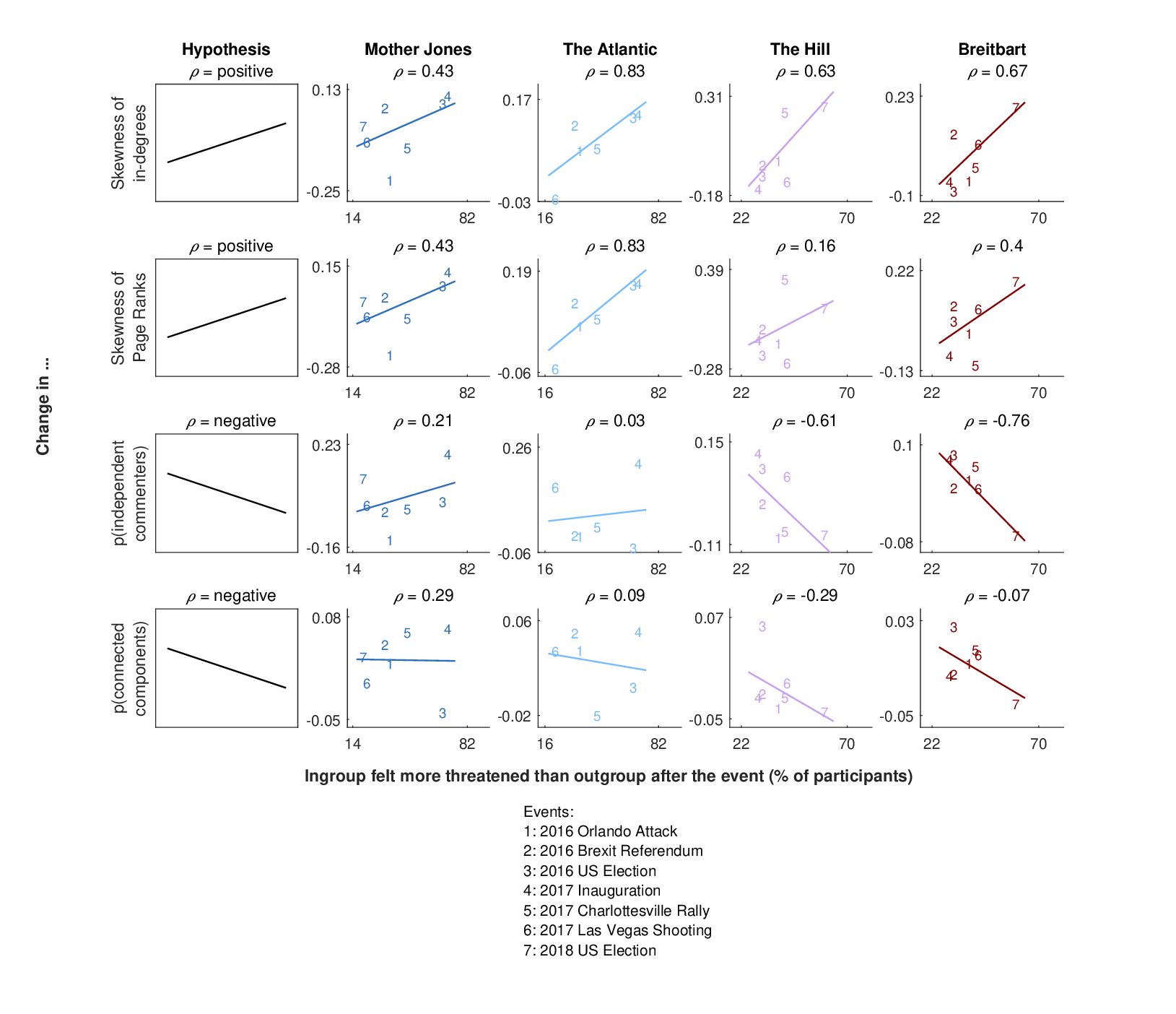}
        \caption{Changes in network measures of inequality \added{of attention} from before to after events rated as more or less threatening to the ingroups on different news sites. The ratings (x-axis) are summarized as the percentage of participants who estimated that the ingroup on a given site (Democrats for Mother Jones and The Atlantic, Republicans for The Hill and Breitbart) felt more threatened by the event than the outgroup (survey question 4, see Figure~\ref{fig:threat_perception}). Rows show results for different measures of inequality of \deleted{influence}\added{attention} (y-axis) on the individual (the first two rows) and the network level (the second two rows). The first column shows the hypothesized median patterns of results, and the other columns show results for different sites. Numbers correspond to different events (see legend), and patterns are summarized by least-squares fitted lines across events and the Spearman $\rho$ correlation coefficient above each plot. Colors denote the political orientation of different news sites: dark blue for the leftmost, Mother Jones, lighter blue for left-leaning The Atlantic, purple for the moderate to right, The Hill, and red for the rightmost, Breitbart.}
     \label{fig:change_threat_ingr2}
\end{figure}

\textbf{The relationship between the inequality of \deleted{commenters' influence}\added{attention} and group threat.} Our first hypothesis is that events perceived as a threat to a collective will be associated with \added[id=r2]{higher collective attention to a few high-profile voices, resulting in} higher levels of inequality of \deleted{influence in}\deleted{attention to different members of}that collective. To investigate this, we compare our four indices of inequality derived \deleted{form}\added{from} commenter networks (skewness of in-degree distributions, skewness of Page Rank distributions, the proportion of independent commenters, and the relative number of connected components) with survey ratings of the level of threat perceived to be associated with different events (see Method for details). As the pattern of survey results in Figure \ref{fig:threat_perception} in Supplementary Material suggests (panels 3-6), events involving a more specific ingroup-outgroup division (2016 and 2018 U.S. Elections, 2017 Inauguration, and 2017 Charlottesville Rally) tended to be \deleted{as}perceived more threatening to the participants' ingroup than events with a less specific outgroup (2016 Orlando Attack, 2016 Brexit Referendum, and 2017 Las Vegas Shooting; see more details in the caption of Figure \ref{fig:threat_perception}).

Figure~\ref{fig:change_threat_ingr2} shows the results of comparing these survey ratings with network-based indices of inequality. Individual measures of inequality (skewness of in-degree and Page Rank distributions, the first two rows in Figure~\ref{fig:change_threat_ingr2}) are positively related to survey ratings of threat to one's ingroup, for both left- and right-leaning sites, as expected. For the network-level measures of inequality (proportion of independent commenters and the relative number of connected components, the second two rows in Figure~\ref{fig:change_threat_ingr2}), the expected negative correlation with ingroup threat is observed for the right-leaning sites, as expected, but not for the left-leaning ones. Correlation coefficients for all trends and their bootstrapped confidence intervals are shown in Table~\ref{tab:correlations}, generally supporting visually observed trends.

Results for the other two survey measures of ingroup threat show the same patterns (Figures~\ref{fig:change_threat_ingr1},~\ref{fig:change_threat_ingr3}). Results for the measures of personal threat (Figures~\ref{fig:change_threat_pers1},~\ref{fig:change_threat_pers2}), and the overall threat to everyone independently of political orientation (Figure~\ref{fig:change_threat_overall}), show similar although less strong patterns, as expected given that those types of threat are not critical for the hypothesized relationship between ingroup threat and \deleted{commenter inequality of influence}\added{inequality of attention to ingroup members}.

\begin{figure}
    \centering
        \includegraphics[width=1\textwidth]{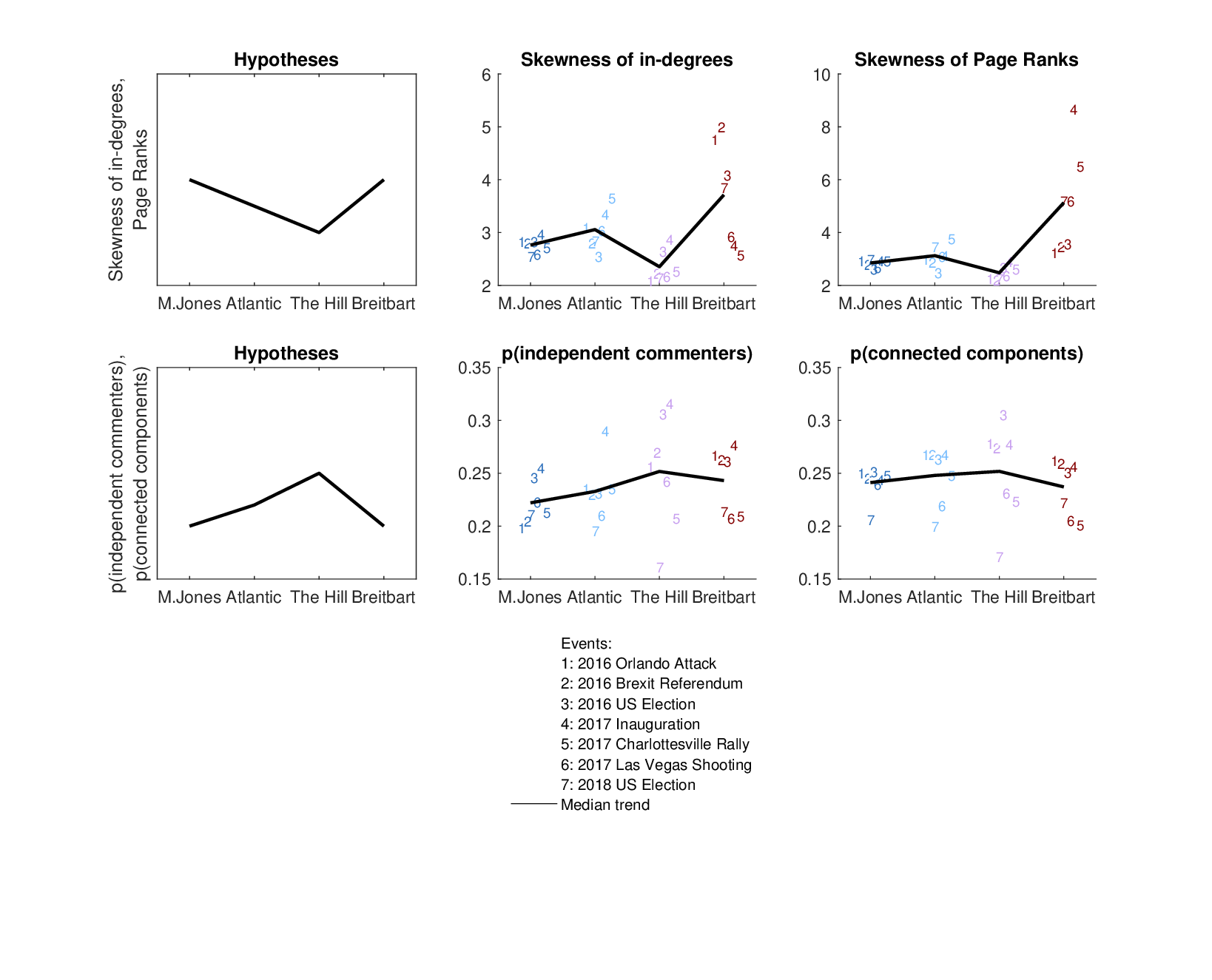}
        \caption{Differences in the overall inequality of \deleted{commenters' influence}\added{attention}, across different news websites and events, measured as the skew of centrality indices (first row) and properties of the overall network structure (second row) for the top 1\% of commenters. The first column shows hypothesized median patterns of results. Numbers correspond to different events (see legend), and the line to the median trend across all events. Colors denote the political orientation of different sites as in Figure~\ref{fig:change_threat_ingr2}.}
     \label{fig:inequal_1p_v2}
\end{figure}

\textbf{The relationship between the inequality of \deleted{commenters' influence}\added{attention} and political extremity.} Our second hypothesis is that commenter networks on politically \deleted{the} more extreme news websites \added[id=r2]{might reflect somewhat higher deference to leaders and show more inequality}\deleted[id=r2]{ will be more unequal in terms} of \deleted{commenters' influence}\added{attention} than networks on more moderate sites. This translates into two hypothesized patterns of results (see the first column in Figure~\ref{fig:inequal_1p_v2}). First, for the individual-level measures of inequality (skewness of individual commenters' centralities) we expect a roughly U-shaped relationship between the political extremity of the four news websites ordered from `very left' to `very right' and the skewness of centrality distributions. Second, for the measures of network-level inequality, we expect an inverted U-shaped relationship between the political extremity and the proportions of independent commenters and connected components. Because The Atlantic is politically more off-center than The Hill is, the difference between Mother Jones and The Atlantic could be smaller than the difference between The Hill and Breitbart, as sketched in the first column in Figure~\ref{fig:inequal_1p_v2}.

As shown in the second and third columns in Figure~\ref{fig:inequal_1p_v2}, there is a quite consistent signal in the data that is roughly in line with the expected patterns, especially for the individual measures of inequality (skewness measures). Specifically, the skewness of distributions of commenters' centralities is larger for the more extreme political sites than for the more centrally positioned one, The Hill (first row in Figure~\ref{fig:inequal_1p_v2}). These patterns are roughly the same no matter whether the skew is calculated only for the top 1\% of commenters (the first row in Figure~\ref{fig:inequal_1p_v2}) or for all commenters (the first row in Figure~\ref{fig:inequal_all_v2}). The only difference is that when the skew is calculated only for the top 1\% of commenters, the skewness of centralities' distributions is particularly large for the far-right site, Breitbart, suggesting higher inequality of \deleted{influence}\added{attention} on that site compared to the left-positioned sites, Mother Jones and The Atlantic. When the skew is calculated for all commenters, this pattern is less clear, possibly because the changes in measures of skew are obscured when calculated on all comments. 

Network-level measures also show slightly more inequality (fewer independent commenters and connected components) for the more extreme left-oriented sites than for the politically more central site, The Hill (the second row in Figure~\ref{fig:inequal_1p_v2}). The far-right site, Breitbart, shows the expected pattern as well, but only for the proportion of connected components, and not for the proportion of independent commenters, where it ties with The Hill. 

Of note, there is variability in all indices of inequality across different events, as visible from the dispersion of the events around the median trend lines in Figure~\ref{fig:inequal_1p_v2}. Therefore, it is important to compare multiple measures to evaluate the robustness of the observed trends. For example, although inequality on Breitbart is larger than on other sites for most events, inequality of \deleted{influence}\added{attention} on Breitbart at the time of the 2017 Inauguration appears larger when measured as skewness of Page ranks than skewness of in-degrees. As another example, although the moderate news site, The Hill, followed expected trends, it had lower than expected proportions of independent commenters and connected components for the 2017 Charlottesville alt-right Rally and the 2018 U.S. election (lower than the more extreme news sites, see events 5 and 7 in Figure~\ref{fig:inequal_1p_v2}). This likely reflects the greater diversity of opinions on this moderate news site, with some events evaluated as in line with a more liberal point of view, and others in line with a more conservative point of view.

\added{\textbf{Further analyses}}

\added{We next conduct several additional analyses to check for possible distributional and temporal confounds in our data. First, we computed the average “age” (days from signing up for Disqus) of commenters participating in discussions before and after different events (first row of Figure \ref{fig:desmes} in the Supplementary Material). We calculated the percentage change in the average age of commenters from before to after each event and analyzed how it relates to the event’s threat rating. If on average participants became ‘younger’ after the event, this suggests that new participants who did not comment before have now joined the discussion. In contrast, if they became ‘older’, this means that the discussion is dominated by more experienced participants.}

\added{We found that on the left-oriented sites commenters become ‘younger’ on average after more threatening events. In contrast, on the right-oriented site Breitbart, commenters become ‘younger’ after least threatening events for their ingroup, such as the 2016 Election and 2017 Inauguration. This provides a hint about possibly different motivations for signing up for discussions on these sites: on the two left-oriented sites the motivation seems to be ‘solving a problem’, or a threat the ingroup is facing, while on the right-oriented site the motivation seems to be more ‘participating in a celebration’ following an event that’s perceived to be a ‘win’ for the group (e.g. the 2016 U.S. Election).}

\added{Critically for our central argument about the relationship of threat and inequality, these age differences cannot explain the relationship of threat and skewness of in-degree and Page Rank distributions shown in the first two rows of Figure \ref{fig:change_threat_ingr2}. While age changes differ across sites, the relationship between threat and skewness goes in the same direction for all four sites. The age change patterns might have, however, contributed to the different patterns of change in the proportion of independent commenters on the left- vs. right-oriented sites (the third row in Figure \ref{fig:change_threat_ingr2}). If the newcomers on the left-oriented sites typically come to help solve the group’s problems, they might have been more likely to offer independent advice than the newcomers on the right-oriented sites who came to celebrate their group’s victories. This might partially explain the slightly positive relationship between threat and the proportion of independent commenters on the left-oriented sites, which was not in line with our hypothesis.}

\added{Second, we analyzed the relationship between the threat ratings and several measures of activity volume before and after each event: the number of comments per article, the number of commenters per article, and the ratio of the number of comments and commenters. These results are shown in the last three rows of Figure \ref{fig:desmes}. While the relationship between threat and the number of comments is not strong and consistent, threat ratings tend to be positively related to the increase in the number of commenters. However, because the number of comments does not increase much or at all, there is a largely negative relationship between threat and the ratio of comments to commenters. This seems to support our overall argument that in times of trouble more people focus on fewer comments, and that these comments are likely contributed by a smaller proportion of very productive commenters.}

\added{Third, we have conducted extensive analyses of the measures of the volume of activity with all measures of inequality, Figure \ref{fig:inequal_articles} in the Supplementary Material shows how different measures of inequality related to the number of articles posted before and after each event. Figure \ref{fig:inequal_comments} shows the same for the number of comments, Figure \ref{fig:inequal_commenters} for the number of commenters, and Figure \ref{fig:inequal_comments_commenters} for the ratio of the number of comments to commenters. The relationships are largely close to zero, except for a couple of relationships of the number of commenters and measures of inequality. First, likely reflecting the patterns uncovered in the age change analysis above, events attracting more commenters are associated with a higher inequality on the left-oriented sites (where new commenters seem to more often come in times of threat) and a lower inequality on the right-oriented sites (where new commenters appear to come when it is time to celebrate ingroup wins). Second, there is a moderately positive relationship between the number of commenters and the proportion of independent commenters. This makes sense as the events attracting more commenters might be attracting people who otherwise comment rarely and have come to make a fast independent comment rather than participating in a discussion. But overall, these analyses show that our indices of inequality are most likely not driven simply by the volume of activity on different sites and at different times.}

\added{Finally, as mentioned before, we have analyzed the frequency with which commenters who posted mostly on one site also commented on the other news sites. While the percentage of such commenters and comments is quite small (see Tables \ref{tab:overlaps1} and \ref{tab:overlaps2}), there are interesting differences between the left- and right-oriented sites. While people who predominantly comment on one of the left-oriented sites (Mother Jones and Atlantic) comment similarly often (though rarely) on each of the other sites (in particular on the moderately oriented site The Hill), people who predominantly comment on the moderate-oriented and right-oriented comment on the left-oriented sites much more rarely than on politically closer sites. That is, commenters on The Hill are much more likely to also comment on Breitbart than on either of the left-oriented sites and similarly, commenters on Breitbart are much more likely to also comment on The Hill than on either Mother Jones or The Atlantic. Although it is difficult to exclude possible adversarial troll activities, these findings are in line with the proposals that the left side of the political spectrum is characterized by a higher openness to diverse experiences \cite{gerber2011big}. More relevant for our main argument on the relationship between threat and inequality, Table \ref{tab:overlaps2} shows that the overall percentage of comments on each site that are posted by commenters who predominantly post on other sites is very small, and therefore unlikely to affect our main conclusions.}

\section*{Discussion}

Using data on commenting patterns of tens of thousands of commenters during a three-year period, across seven different important societal events and four large news sites, we study changes in the distribution of \deleted{influence}\added{attention} within groups supporting different political options and experiencing real or imagined threats. Unlike in controlled laboratory experiments of network change \cite{almaatouq2020adaptive}, we could use only noisy observational measures of some aspects of relevant group behavior. However, focusing on trends that emerge across seven different societal contexts and individual- and group-level measures of inequality, we observe patterns that are in general agreement with theoretical expectations. 

First, we find that groups experiencing real or imagined threat from specific other groups tend to show an increase in inequality of \deleted{influence}\added{attention}, whereas groups experiencing lower threat tend to show a decrease in inequality (Figure \ref{fig:change_threat_ingr2}). For example, after the 2016 U.S. Election in which Trump won over Clinton, leading to higher ratings of ingroup threat on the left side of the political spectrum, commenters on the left-leaning websites, Mother Jones and The Atlantic, tend to show higher inequality, while commenters on the right-leaning websites, Breitbart and The Hill, tend to show patterns of lower inequality. Similarly, after the 2018 U.S. Election, when Democrats gained control in the House of Representatives, increasing the perceived ingroup threat on the right side of the political spectrum,  commenters on the right-leaning websites show patterns of higher inequality of \deleted{influence}\added{attention}, while commenters on the left-leaning website, Mother Jones, show patterns of lower inequality. These patterns are in line with previous theoretical and empirical observations about the effects of outgroup threat on group functioning \cite{janis1982groupthink,turner1992threat}. 
They are also in line with sociological accounts of how groups manage uncertainty. In situations of heightened cultural uncertainty that might ensue after critical societal events like those studied here, groups might choose to decrease their social uncertainty by heightening the normative influence of a smaller number of individuals \cite{white2008identity}. Similar trade-offs between cultural and social uncertainty and control have been hypothesized in studies of suicide \cite{durkheim1952study} and societal rituals \cite{douglas1970natural}. 

Second, we find that commenters on the more extreme political sites tend to show more inequality of \deleted{influence}\added{attention} than commenters on the more moderate political sites (Figure~\ref{fig:inequal_1p_v2}). The extreme-right site, Breitbart, shows particularly strong inequality of \deleted{influence}\added{attention} measured as skewness of in-degrees and Page Ranks, although only among the top 1\% of commenters (Figure~\ref{fig:inequal_1p_v2} vs. Figure~\ref{fig:inequal_all_v2}). These findings are in line with the past evidence about the heightened deference to authorities among radical proponents on both sides \cite{shils1954authoritarianism,de2011left}, but also provide some support to the theoretical perspectives linking authoritarianism with right-wing political extremity \cite{jost2003political}. 

Studies of network dynamics have typically investigated the influence of diverse socio-psychological factors in controlled laboratory settings \cite{almaatouq2020adaptive,harrell2018strength,rand2011dynamic}. The abundance of online discourse provides a convenient opportunity for studying such factors over tens of thousands of commenters and long periods of time. Analysis of online discussions can yield valuable insights about real-world group dynamics, with many people seeing these discussions as similar to those they have in other platforms and places \cite{duggan2016political}, and beliefs and sentiments formed online spilling over into the ``real'' life \cite{wapo2021riot}.

An advantage of this study is that it overcomes many of the obvious algorithmic reasons for inequalities in commenters' \deleted{influence}\added{attention}. Social media platforms employ a variety of algorithms designed to promote content that is more engaging and profitable, and these technologies differ from site to site. \added{Because all four news sites we investigated used the same commenting platform, Disqus, we were able to compare the effects of the socio-psychological factors that influence discussions without such technological confounds. Besides removing possible effects of different commenting interfaces, Disqus is suitable for research because it does not employ any algorithms that personalize the content users see in line with their preferences. As described before, it enables users to change the default sorting order and the moderators to implement different degrees of automated moderation, but these uncontrollable differences are unlikely to introduce systematic biases that would mimic the intricate patterns of the relationship between threat and inequality we found. Our further analyses of possible confounding effects of new users and volume of activity suggest that none of these effects can explain the whole pattern of our results.} Further research could also investigate shorter or longer-term changes in commenters' networks. In this study, we focused on the one month before and after a critical event as this provided us with a sufficient amount of data from all sites. A longer time would have been complicated in the studied period (2016-2018) as many other societally significant events occurred on at least a monthly basis, and with longer time scales, it would become increasingly more difficult to account for them. 

Studying the behavior of an adaptive complex social system such as commenter networks on political news websites is also extremely difficult. Many factors beyond those that we measured influence commenters' behavior, and it is not easy to determine the mechanisms underlying some users' higher popularity. \added[id=r2]{There are probably several aspects of the internal communication dynamics in the comment sections that contribute to consistently more replies received by some individuals. One is that these individuals might be those who comment earlier in the discussion. Another is that some commenters are actually more popular and the other users are more likely to reply to them when they see their comment (even if they do not necessarily search for them), further reinforcing their popularity. During our extensive qualitative research of these sites, we noticed several such popular commenters who often got a lot of replies and were referenced in discussions even when they did not comment themselves. A third mechanism is that – as also glimpsed from our qualitative research - commenters quite often check and comment on others’ profiles and specifically on how many comments they have already posted. Commenters with just a few comments run the risk of being accused of trolling, while those with many comments might be taken more seriously and ‘deserving’ of a reply. A fourth reason is discourse dynamics, that is specific topics and phrases that particular commenters introduce and use.} Further qualitative and large-scale natural language processing studies would help better understand the mechanisms driving the inequalities we observe. In particular, it would allow one to investigate whether people who attract the most comments spread their ideas to others\added{ or have their ideas criticized, whether they} harass others or are harassed, \deleted{or}\added{and whether they} show other differences in their discourse compared to commenters who attract fewer replies. \added[id=r2]{Importantly, independently of the mechanism by which commenters choose to whom to reply, they can always choose to post an independent comment. The latter choice will affect the inequality of attention to different participants in the discussion.}

This study contributes to the broader literature on collective adaptation to societal events, including collective attention (\cite{kobayashi2021modeling}, emotion \cite{goldenberg2020collective}, and memory \cite{hirst2018collective}). In this large-scale investigation of real-world interactions, we show that collectives adapt to \added{the} real or imagined threat from an outgroup by changing their network structure, \added{paying more attention to fewer group members and thus potentially allowing them to have} \deleted{enabling }more influence \deleted{of fewer ingroup members}. This can be helpful in the short run to organize a coordinated defense \cite{kirke2010military} but can have adverse consequences in the long run, such as a lower ability to consider diverse views when solving complex problems \cite{janis1982groupthink}. While we used data from a particular type of discourse, given the wide range of threatening events and levels of political extremity we studied, we believe these results are roughly indicative of group reactions in other public discussion contexts.

\section*{Data Availability} All of the comments data is freely available to download through Disqus API \cite{disqus}. The data from survey results are available from the corresponding author upon reasonable request.  

\section*{Acknowledgement}
This work was supported by a grant from the National Science Foundation (DRMS 1757211). The funder had no role in the study design or interpretation of results. We thank Joshua Garland, Haiko Lietz, Allison Morgan, Henrik Olsson, and Kenan Turbic for their helpful comments.

\section*{Author Contributions} N.G.B.T.* collected and analyzed the data, and N.G.B.T.* and M.G. contributed to presenting the figures and writing the paper. Both authors reviewed the manuscript.

\section*{Competing Interests} The Authors declare that there is no conflict of interest. 

\section*{Materials \& Correspondence} N.G.B.T.* is the author for correspondence and materials requests. 

\bibliographystyle{naturemag}
\bibliography{ref} 
\clearpage

\renewcommand \thesection{S}
\pagenumbering{gobble}

\section*{N. Gizem Bacaksizlar Turbic and Mirta Galesic: Group Threat, Political Extremity, and Collective Dynamics in Online Discussions\\\\
Supplementary Material}

\renewcommand{\thetable}{\Alph{section}\arabic{table}}
\renewcommand\thefigure{\thesection\arabic{figure}}    
\setcounter{figure}{0}  
\setcounter{table}{0}  
\renewcommand{\bibnumfmt}[1]{[S#1]}
\renewcommand{\citenumfont}[1]{S#1}
\renewcommand{\thetable}{S\arabic{table}}
\renewcommand{\thefigure}{S\arabic{figure}}

\begin{table} [hbt!]
	\centering
	\caption{Commenter, comment, and article counts for articles published by Mother Jones, The Atlantic, The Hill, and Breitbart, during the month before and the month after different events.}
	\resizebox{\textwidth}{!}{%
	\begin{tabular}{|c||c||c||r|r|r||r|r|r|}
	    \cline{4-9} 
	    \multicolumn{3}{c|}{} &
	    \multicolumn{3}{c||}{Month Before} & \multicolumn{3}{c|}{Month After} \\
		\hline
		\multicolumn{1}{|c||}{\textbf{Event}} & \multicolumn{1}{|c||}{\textbf{Date}} & \multicolumn{1}{c||}{\textbf{News Site}} & \textbf{Commenters} & \textbf{Comments} & \textbf{Articles}& \textbf{Commenters} & \textbf{Comments} & \textbf{Articles} \\ 
		\hline\hline
		& \multirow{4}{*}{12-Jun-2016} & Mother Jones & 7,336 & 95,862 &912 & 5,456 & 86,830 & 887\\
		\cline{3-9}
		1. Orlando &  & The Atlantic & 13,338 & 185,918 & 2,153 & 12,359 & 181,401 & 2,217\\ 
		\cline{3-9}
		Attack & & The Hill & 35,203 & 1,480,553 & 4,759 & 31,596 & 1,523,329 & 4,743\\
		\cline{3-9}
		& & Breitbart & 46,966 & 2,017,814 & 8,283 & 50,800 & 2,052,007 & 8,077 \\ 
		\hline\hline
		   \multirow {4}{*}{2. Brexit} & \multirow {4}{*}{23-Jun-2016} & Mother Jones & 5,948 & 93,130 & 468 & 5,103 & 77,151 & 453\\
		\cline{3-9}
		  & & The Atlantic & 13,223 & 188,074 & 1,320 & 11,829 & 171,601 & 1,277 \\ 
		\cline{3-9}
		 Referendum & & The Hill & 33,595 & 1,538,196 & 3,697 & 34,414 & 1,574,227 & 3,436 \\ 
		\cline{3-9}
		& & Breitbart & 48,695 & 2,017,959 & 20,956 & 52,226 & 2,231,633 & 27,917	\\ 
		\hline\hline
		& \multirow{4}{*}{8-Nov-2016} & Mother Jones & 4,517 & 251,042 & 2,137 & 5,021 & 249,643 & 1,956\\
		\cline{3-9}
		3. United States &  & The Atlantic & 10,738 & 619,849 & 5,152 & 9,723 & 517,866 & 4,254\\ 
		\cline{3-9}
		Election & & The Hill & 35,474 & 1,818,410 & 5,732 & 37,201 & 4,459,860 & 13,713\\ 
		\cline{3-9}
		& & Breitbart & 59,449 & 7,946,536 & 22,058& 66,565 & 8,774,603 & 24,405 \\ 
		\hline\hline
		& \multirow{4}{*}{20-Jan-2017} & Mother Jones & 4,255 & 75,229 & 730 & 5,420 & 86,183 & 863\\
		\cline{3-9}
		4. Presidential &  & The Atlantic & 8,974 & 156,660 & 1,751 & 12,050 & 194,933 & 1,908 \\ 
		\cline{3-9}
		Inauguration & & The Hill & 30,681 & 1,536,162 & 4,877 & 41,080 & 1,974,873 & 5,883\\ 
		\cline{3-9}
		& & Breitbart & 61,052 & 2,701,699 & 9,199 & 72,863 & 3,495,675 & 10,159 \\ 
		\hline\hline
		& \multirow{4}{*}{11-Aug-2017} & Mother Jones & 3,853 & 84,214 & 875 & 3,316 & 77,417 & 780\\
		\cline{3-9}
		5. Charlottesville & & The Atlantic & 10,147 & 215,257 & 2,071 &
		10,915 & 228,683 & 1,839 \\ 
		\cline{3-9}
		Rally & & The Hill & 36,147 & 2,091,593 & 6,110 & 32,557 & 1,992,493 & 5,460 \\ 
		\cline{3-9}
		& & Breitbart & 54,180 & 3,415,882 & 8,824 & 60,607 & 3,791,626 & 9,009 \\ 
		\hline\hline
		& \multirow{4}{*}{1-Oct-2017} & Mother Jones & 3,085 & 65,009 & 771& 3,225 & 64,423 & 801 \\
		\cline{3-9}
		6. Las Vegas &  & The Atlantic & 7,860 & 174,900 & 1,873 & 9,147 & 218,059 & 1,870\\ 
		\cline{3-9}
		Shooting & & The Hill & 33,258 & 1,826,011 & 5,949 & 38,152 & 2,278,911 & 6,507\\ 
		\cline{3-9}
		& & Breitbart & 57,811 & 3,541,065 & 9,072&60,258 & 3,817,425 & 9,458\\ 
		\hline\hline
		& \multirow{4}{*}{6-Nov-2018} & Mother Jones & 2,343  & 199,326 &1,935 & 2,697  & 175,379 & 1,840\\
		\cline{3-9}
		7. United States &  & The Atlantic & - & - & - &- & - & -\\
		\cline{3-9}
		Election & & The Hill & 30,285 & 7,902,881 & 14,733 & 27,753 & 8,058,085 & 15,295\\ 
		\cline{3-9}
		& & Breitbart & 58,928 & 10,613,517 & 23,867 & 55,647 & 8,842,971 & 19,236 \\ 
		\hline
	\end{tabular}
	\label{tab:counts}
    \vspace{-0.4cm}
   }
\end{table}

\begin{table}[hbt!]
\centering
    \caption{The top five events from the U.S. news according to Google Trends from 2016 - 2018 \cite{trends}.}
    \label{tab:events}
    \begin{tabular}{|l|c|c|c|}
    \hline
        Rank & 2016 U.S. News  & 2017 U.S. News  & 2018 U.S. News  \\ \hline
        1 & Olympics & Hurricane Irma & World Cup  \\ \hline
        2 & Election & Las Vegas Shooting  & Hurricane Florence  \\ \hline
        3 & Orlando Shooting & Solar Eclipse  & Mega Millions \\ \hline
        4 & Brexit & Hurricane Harvey  & Election Results \\ \hline
        5 & Zika Virus & Bitcoin Price  & Hurricane Michael \\ \hline
    \end{tabular}
    
\end{table}

\begin{sidewaystable}[hbt!]

\begin{adjustwidth}{-2.5 cm}{-2.5 cm}\centering\begin{threeparttable}[!htb]
    \caption{Correlation coefficients and bootstrapped confidence intervals for trends shown in Figures 2, A5-A9, jointly for the two left-leaning sites (Mother Jones and The Atlantic) and the  moderate and right-leaning sites (The Hill and Breitbart).}
    \label{tab:correlations}
    \scriptsize
    \begin{tabular}{lrrrrr}\toprule
    &\multicolumn{4}{c}{\textbf{Correlation of survey ratings of threat and network indices of:}} \\\midrule
    \textbf{} &\textbf{Skewness of in-degree} &\textbf{Skewness of Page Ranks} &\textbf{p(independent commenters)} &\textbf{p(connected components)} \\
    \textbf{} &\textbf{} &\textbf{} &\textbf{} &\textbf{} \\
    &\multicolumn{4}{c}{\textbf{Left-oriented news sites (Mother Jones and The Atlantic)}} \\
    Ingroup hurt more than, or same as, outgroup after the event &0.72 [0.31, 0.87] &0.76 [0.39, 0.89] &0.3 [-0.41, 0.77] &0 [-0.66, 0.61] \\
    Ingroup felt more threatened than outgroup after the event &0.55 [0.09, 0.82] &0.62 [0.08, 0.85] &0.23 [-0.57, 0.75] &-0.06 [-0.72, 0.56] \\
    Ingroup felt more threatened after than before the event &0.54 [0.11, 0.78] &0.61 [0.14, 0.83] &0.33 [-0.48, 0.83] &-0.07 [-0.75, 0.55] \\
    Personal feeling of threat after the event &0.2 [-0.48, 0.62] &0.29 [-0.41, 0.67] &0.33 [-0.26, 0.71] &-0.23 [-0.74, 0.3] \\
    Personally felt more threatened after than before the event &0.36 [-0.26, 0.69] &0.43 [-0.26, 0.75] &0.23 [-0.47, 0.72] &-0.19 [-0.75, 0.39] \\
    Overall, people felt more threatened after than before the event &0.24 [-0.33, 0.61] &0.31 [-0.25, 0.63] &0.2 [-0.38, 0.6] &-0.39 [-0.82, 0.11] \\
    & & & & \\
    &\multicolumn{4}{c}{\textbf{Moderate- and right-oriented news sites (The Hill and Breitbart)}} \\
    Ingroup hurt more than, or same as, outgroup after the event &0.78 [0.5, 0.95] &0.5 [0.08, 0.79] &-0.76 [-0.92, -0.47] &-0.6 [-0.81, -0.05] \\
    Ingroup felt more threatened than outgroup after the event &0.7 [0.2, 0.95] &0.41 [-0.18, 0.75] &-0.65 [-0.9, -0.28] &-0.45 [-0.75, 0.06] \\
    Ingroup felt more threatened after than before the event &0.57 [-0.04, 0.88] &0.29 [-0.37, 0.69] &-0.59 [-0.84, -0.17] &-0.42 [-0.76, 0.07] \\
    Personal feeling of threat after the event &0.14 [-0.41, 0.51] &-0.03 [-0.63, 0.42] &-0.3 [-0.68, 0.16] &-0.1 [-0.53, 0.36] \\
    Personally felt more threatened after than before the event &0.27 [-0.39, 0.62] &0.06 [-0.56, 0.55] &-0.43 [-0.79, 0.2] &-0.32 [-0.78, 0.36] \\
    Overall, people felt more threatened after than before the event &0.39 [-0.24, 0.73] &0.17 [-0.54, 0.63] &-0.45 [-0.8, 0.1] &-0.13 [-0.61, 0.39] \\
    \bottomrule
    \end{tabular}
    \end{threeparttable}\end{adjustwidth}
    
\end{sidewaystable}

\begin{table} [hbt!]
\centering
\caption{\added{Percentage of commenters on each site who commented on other sites at least once.}}
\label{tab:overlaps1} 
\begin{tabular}{l|rrrr} \hline
    \makecell[l]{Of all \\commenters who comment \\ mostly \\ on ..} & \multicolumn{4}{l}{… percentage who also commented on other site at least once:}\\
    &Mother Jones & The Atlantic & The Hill & Breitbart \\ \hline                                     
    Mother Jones & 100\% & 13\% & 13\% & 9\% \\
    The Atlantic & 5\% & 100\% & 8\% & 6\%   \\
    The Hill & 8\% & 10\% & 100\% & 28\% \\
    Breitbart & 4\% & 5\% & 18\% & 100\% \\
    \hline
\end{tabular}   
\end{table}

\begin{table} [hbt!]

    \centering
    \caption{\added{Percentage of comments on other sites, posted by commenters on each site.}}
    \label{tab:overlaps2} 
    \begin{threeparttable}
    \begin{tabular}{l|rrrr} \hline
        \makecell[l]{Of all \\commenters who comment \\ mostly \\ on ..} & \multicolumn{4}{l}{… percentage who also commented on other site at least once:}\\
    &Mother Jones & The Atlantic & The Hill & Breitbart \\ \hline                                          
    Mother Jones & 91\% & 4\% & 3\% & 2\% \\
    The Atlantic & 1\% & 95\% & 2\% & 2\%   \\
    The Hill & 1\% & 1\% & 90\% & 8\% \\
    Breitbart & 0\% & 1\% & 3\% & 96\% \\
    \hline
    \end{tabular} 
    \begin{tablenotes}
            \item Note: The main site of a commenter was determined as the site at which they posted most comments.
        \end{tablenotes}
        \end{threeparttable}
\end{table}

\begin{sidewaysfigure}[ht]
    \centering
        \includegraphics[width=1\textwidth]{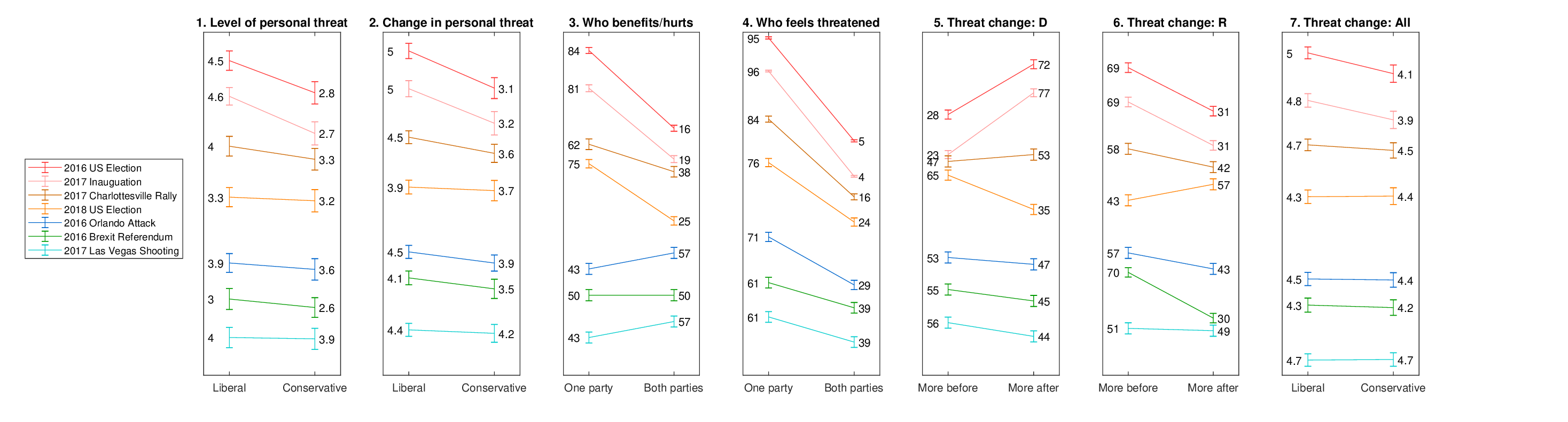}
        \caption{Perceived threat of societal events to different groups, for events hypothesized to be specific group threats (warm colors) and unspecific group threats (cold colors). See details in Method: Societal events. Questions in each panel were: 1. How threatened did you personally feel in the month after this event? \textit{(1-Not at all threatened...7-Extremely threatened)}, 2. In the month after the event, did you feel more or less threatened compared to the month before the event? \textit{(1-I felt much less threatened...7-I felt much more threatened)}, 3. In your view, did the event clearly benefit Democrats at the expense of Republicans, Republicans at the expense of Democrats, or were both groups affected similarly? \textit{(The event benefited Democrats at the expense of Republicans; The event benefited Republicans at the expense of Democrats; The event benefited both Democrats and Republicans; The event hurt both Democrats and Republicans)}, 4. Who do you think felt more threatened in the month after this event, Democrats or Republicans? \textit{(1-Democrats felt much more threatened...7-Republicans felt much more threatened)}, 5. Thinking now of Democrats, do you think they felt more or less threatened in the month after the event compared to the month before this event? \textit{(1-Democrats felt much less threatened...7-Democrats felt much more threatened)}, 6. Thinking now of Republicans, do you think they felt more or less threatened in the month after the event compared to the month before this event? \textit{(1-Republicans felt much less threatened...7-Republicans felt much more threatened)}, 7. Thinking now of all people in  the U.S., independently of their political extremity: In the month after the event, did most people feel more or less threatened compared to the month before the event? \textit{(1-Most people felt much less threatened...7-Most people felt much more threatened)}. \added{Confidence intervals comprise on average +/- 4\% of the mean of each rating, suggesting a strong agreement among participants about how threatened different groups should be - perhaps not surprising in today's extremely polarized political climate.}
        }
    \label{fig:threat_perception}
\end{sidewaysfigure}

\begin{figure}[ht]
    \centering
        \includegraphics[width=.4\textwidth]{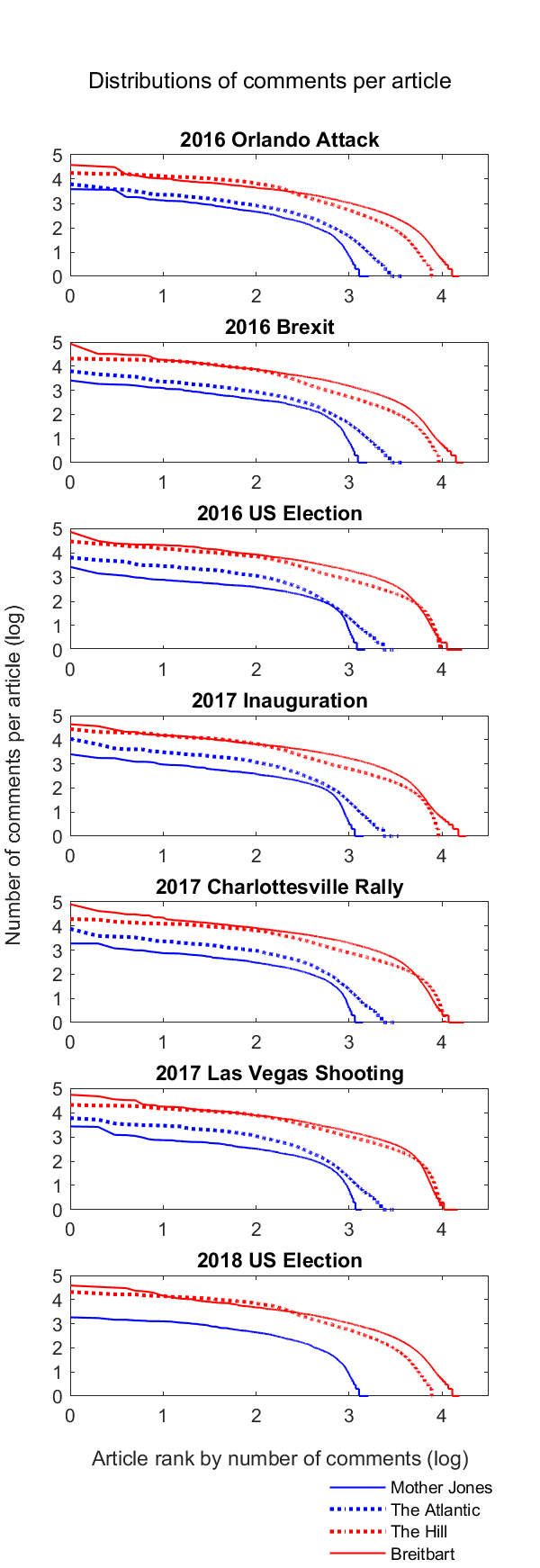}
        \caption{Distributions of the number of comments per article for different websites and events.}
    \label{fig:comart_distributions}
\end{figure}

\begin{figure}[ht]
    \centering
        \includegraphics[width=.7\textwidth]{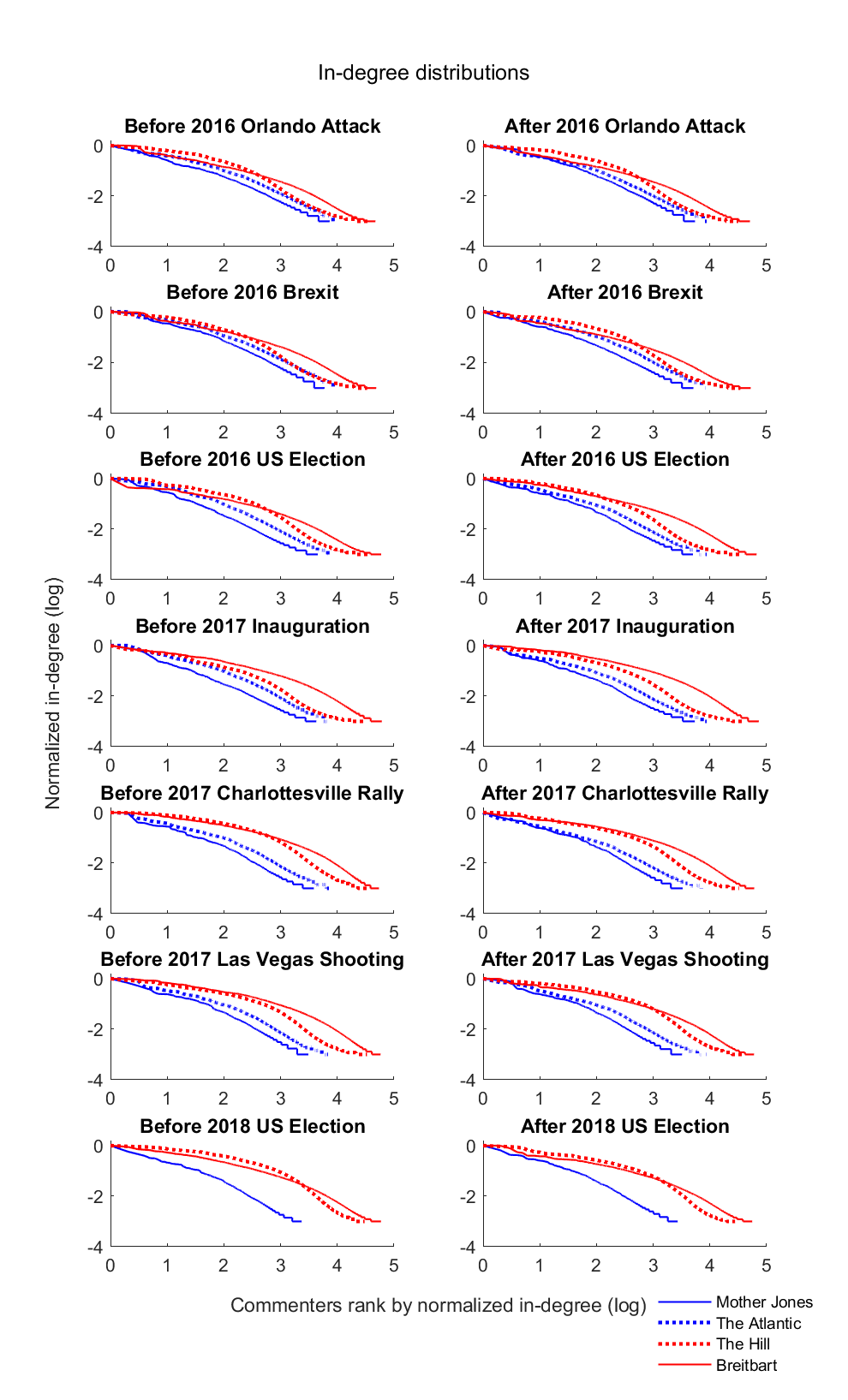}
        \caption{Distributions of commenters' in-degrees before and after important events.}
    \label{fig:indegree_distributions}
\end{figure}

\begin{figure}[ht]
    \centering
        \includegraphics[width=.7\textwidth]{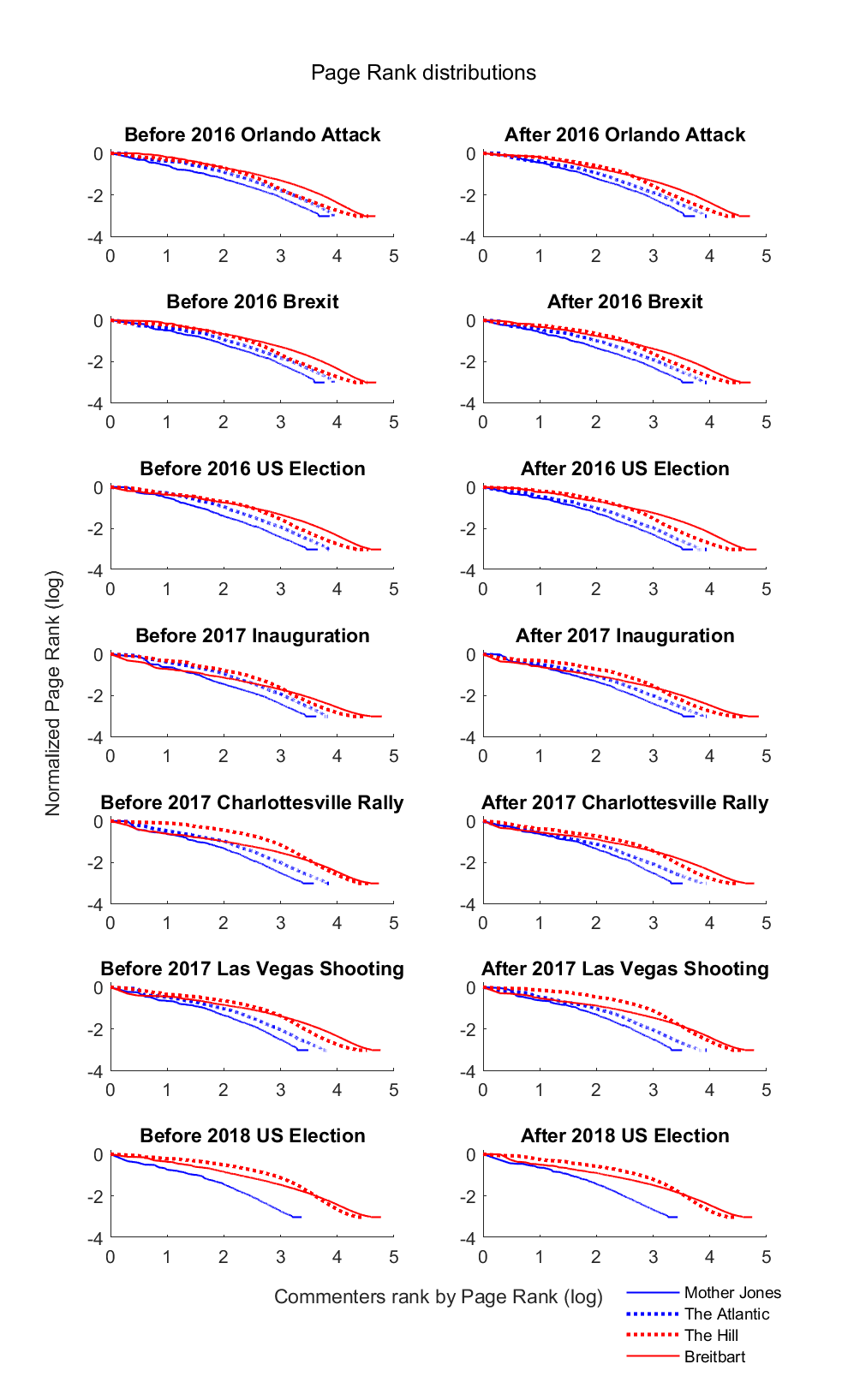}
        \caption{Distributions of commenters' Page Ranks before and after important events.}
    \label{fig:pagerank_distributions}
\end{figure}

\begin{figure}[ht]
    \centering
        \includegraphics[width=1\textwidth]{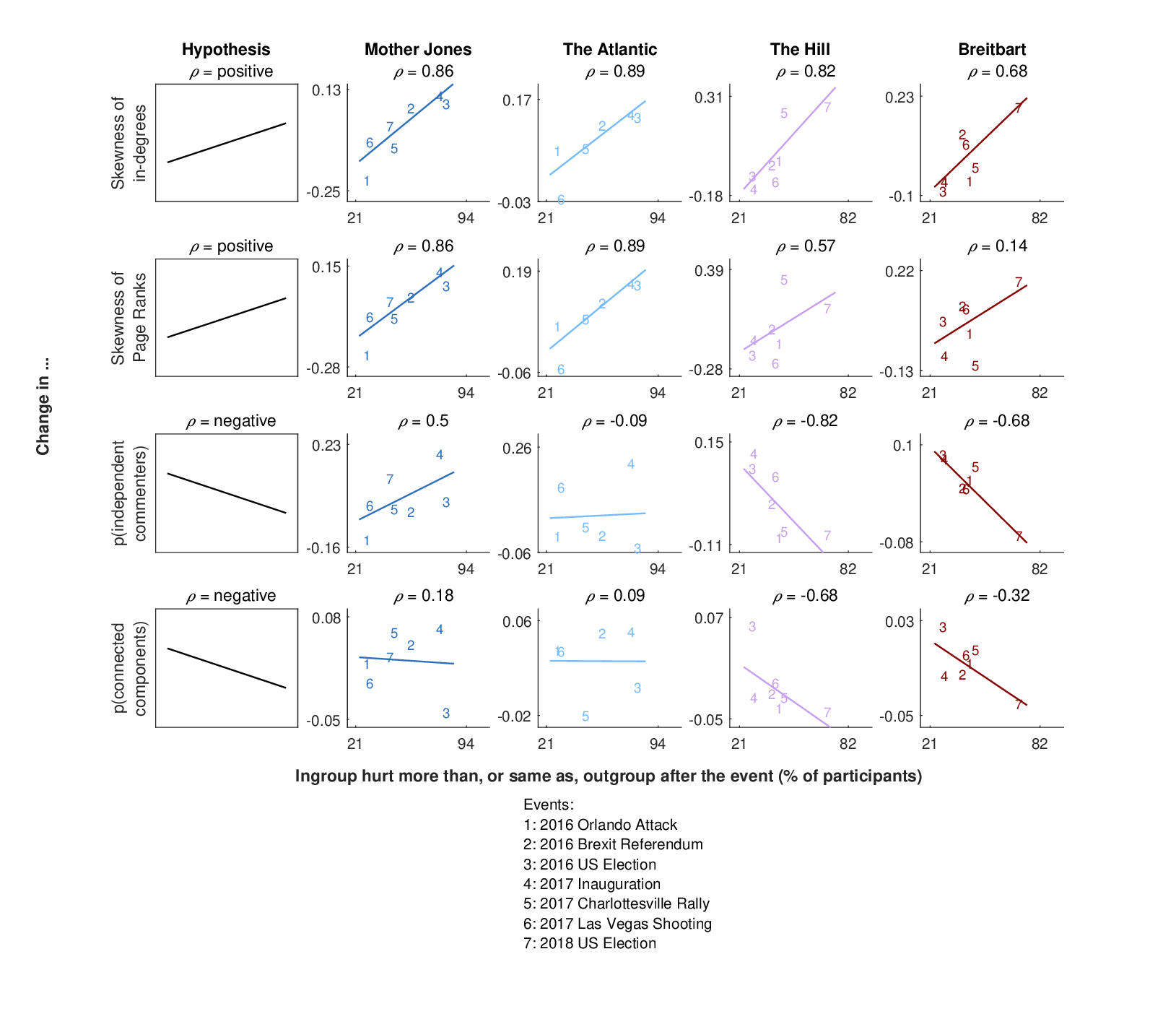}
        \caption{Changes in network measures of inequality from before to after events rated as more or less threatening to the ingroups on different news sites. The ratings (x-axis) are summarized as the percentage of participants who estimated that the event hurt the ingroup (Democrats for Mother Jones and The Atlantic, Republicans for The Hill and Breitbart) the same or more than the outgroup (survey question 3, see Figure~\ref{fig:threat_perception}). Rows show results for different measures of inequality of \deleted{influence}\added{attention} (y-axis), on the individual (the first two rows) and the network level (the second two rows). The first column shows the hypothesized median patterns of results, and the other columns show results for different sites. Colored numbers in each plot correspond to different events (see legend), and patterns are summarized by least-squares fitted lines across events and the Spearman $\rho$ correlation coefficient above each plot.}
     \label{fig:change_threat_ingr1}
\end{figure}

\begin{figure}[ht]
    \centering
        \includegraphics[width=1\textwidth]{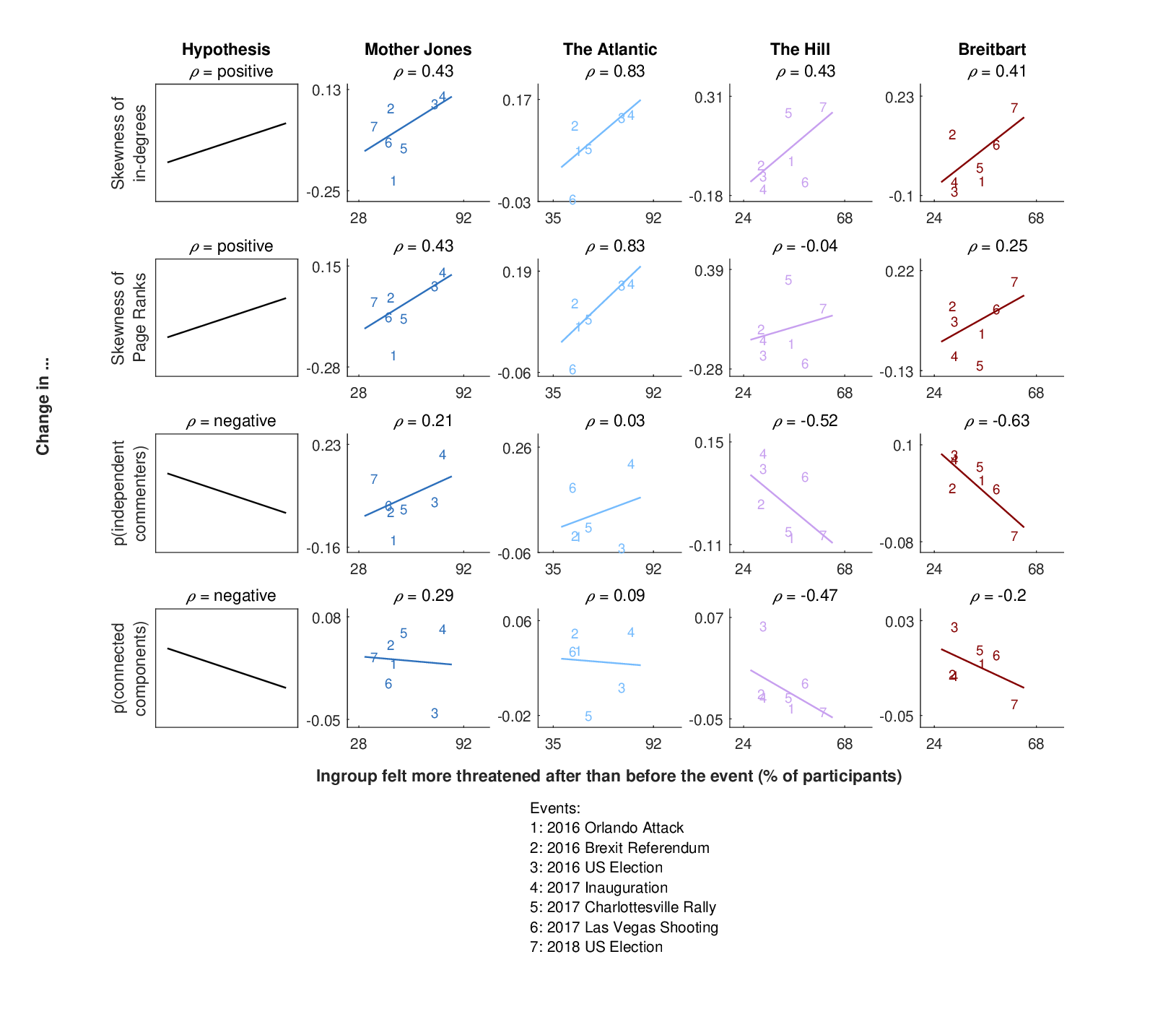}
        \caption{Changes in network measures of inequality from before to after events rated as more or less threatening to the ingroups on different news sites. The ratings (x-axis) are summarized as the percentage of participants who estimated that the ingroup (Democrats for Mother Jones and Atlantic, Republicans for The Hill and Breitbart) felt more threatened in the month after the event compared to the month before the event (5 or more on the 7-point scale, see Figure~\ref{fig:threat_perception}, survey questions 5 and 6). Rows show results for different measures of inequality of \deleted{influence}\added{attention} (y-axis), on the individual (the first two rows) and the network level (the second two rows). The first column shows the hypothesized median patterns of results, and the other columns show results for different sites. Colored numbers in each plot correspond to different events (see legend), and patterns are summarized by least-squares fitted lines across events and the Spearman $\rho$ correlation coefficient above each plot.}
     \label{fig:change_threat_ingr3}
\end{figure}

\begin{figure}[ht]
    \centering
        \includegraphics[width=1\textwidth]{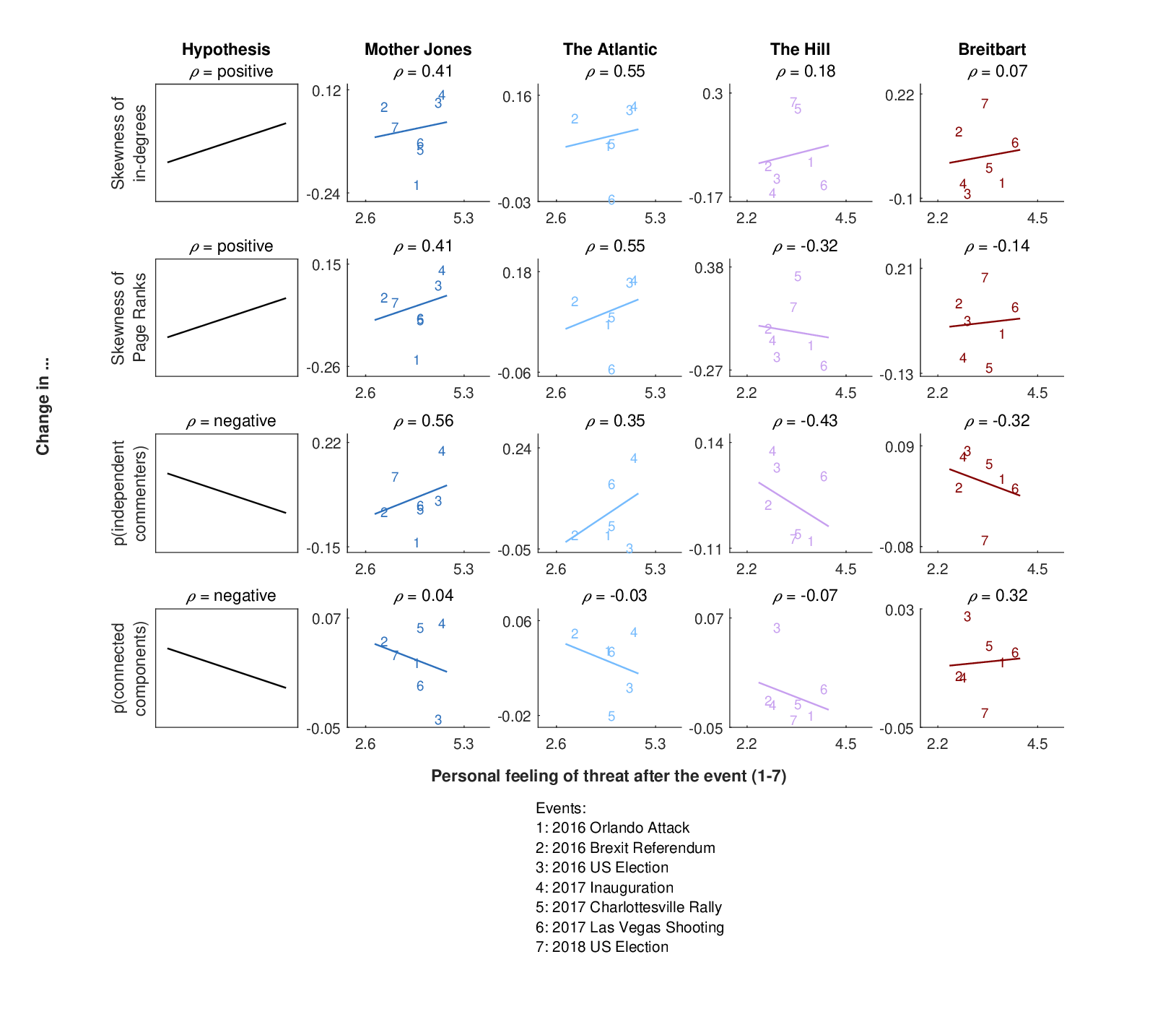}
        \caption{Changes in network measures of inequality from before to after events rated as personally more or less threatening to our liberal (for Mother Jones and The Atlantic) or conservative participants (for The Hill and Breitbart). The ratings (x-axis) are on a scale from 1 to 7, where 7 is the most threat (see Figure~\ref{fig:threat_perception}, survey question 1). Rows show results for different measures of inequality of \deleted{influence}\added{attention} (y-axis) on the individual (the first two rows) and the network level (the second two rows). The first column shows the hypothesized median patterns of results, and the other columns show results for different sites. Colored numbers in each plot correspond to different events (see legend), and patterns are summarized by least-squares lines across events and the Spearman $\rho$ correlation coefficient above each plot.}
     \label{fig:change_threat_pers1}
\end{figure}

\begin{figure}[ht]
    \centering
        \includegraphics[width=1\textwidth]{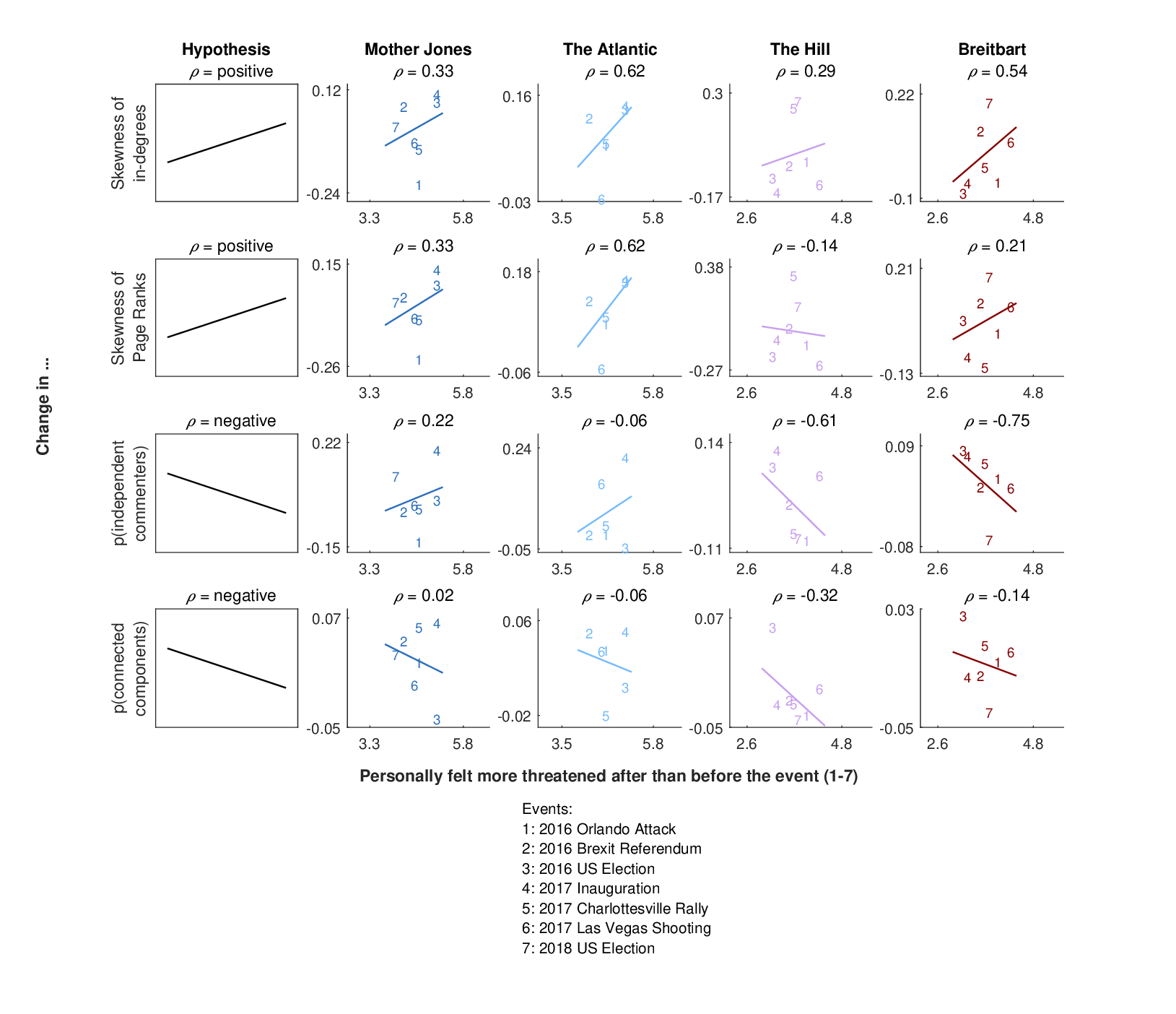}
        \caption{Changes in network measures of inequality from before to after events for which our liberal (for Mother Jones and Atlantic) or conservative participants (for The Hill and Breitbart) experienced more or less threat after compared to before the event. The ratings (x-axis) are on a scale from 1 to 7, where 7 is more threat after the event (see Figure~\ref{fig:threat_perception}, survey question 2). Rows show results for different measures of inequality of \deleted{influence}\added{attention} (y-axis) on the individual (the first two rows) and the network level (the second two rows). The first column shows the hypothesized median patterns of results, and the other columns show results for different sites. Colored numbers in each plot correspond to different events (see legend), and patterns are summarized by least-squares fitted lines across events and the Spearman $\rho$ value above each plot.}
     \label{fig:change_threat_pers2}
\end{figure}

\begin{figure}[ht]
    \centering
        \includegraphics[width=1\textwidth]{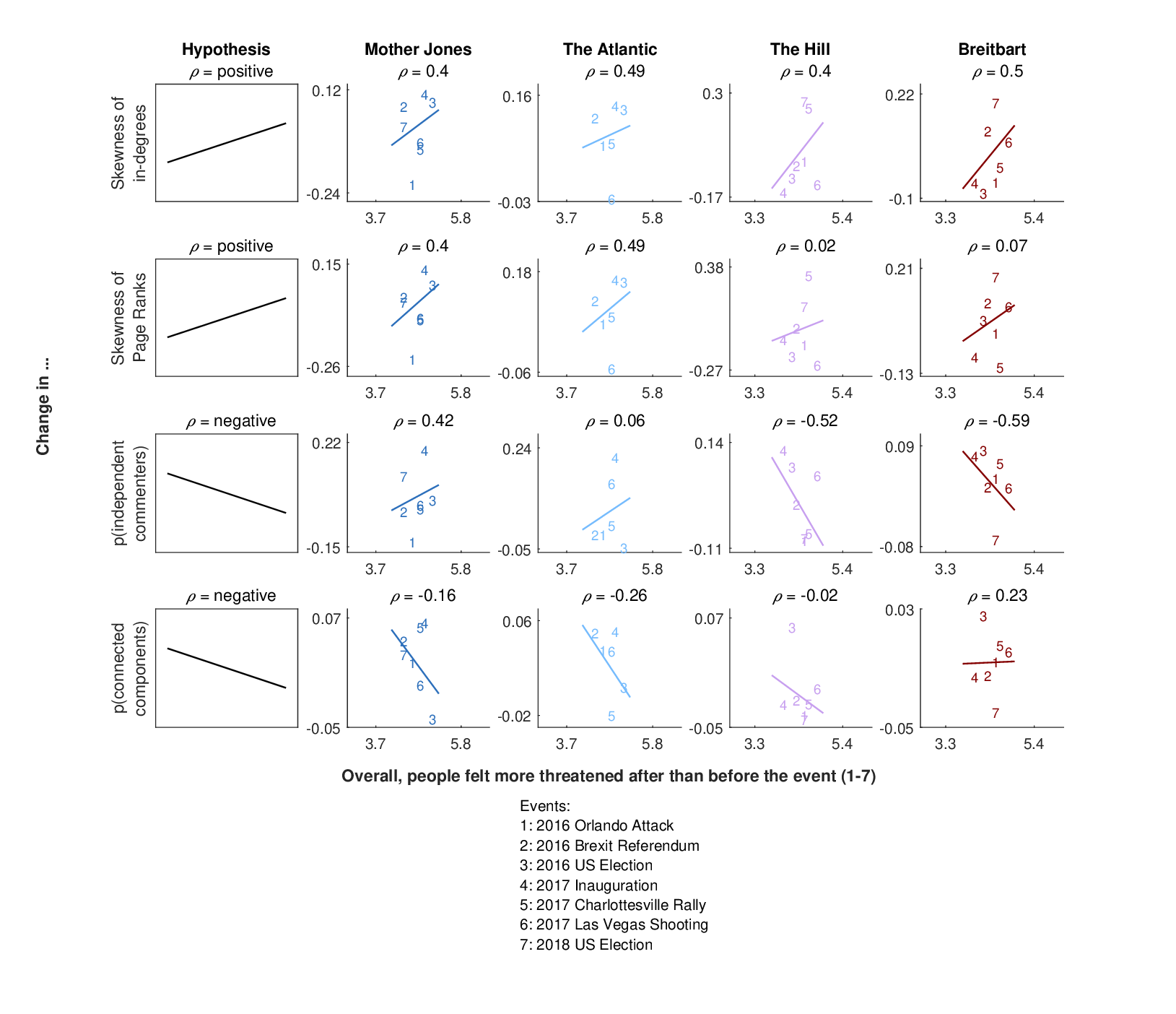}
        \caption{Changes in network measures of inequality from before to after events rated as more or less threatening to most people. The ratings (x-axis) are on a scale from 1 to 7, where 7 is more threat after the event for most people (see Figure~\ref{fig:threat_perception}, survey question 7). Rows show results for different measures of inequality of \deleted{influence}\added{attention} (y-axis), on the individual (the first two rows) and the network level (the second two rows). The first column shows the hypothesized median patterns of results, and the other columns show results for different sites. Colored numbers in each plot correspond to different events (see legend), and patterns are summarized by least-squares fitted lines across events and the Spearman $\rho$ correlation coefficient above each plot.}
     \label{fig:change_threat_overall}
\end{figure}

\begin{figure}[ht]
    \centering
        \includegraphics[width=1\textwidth]{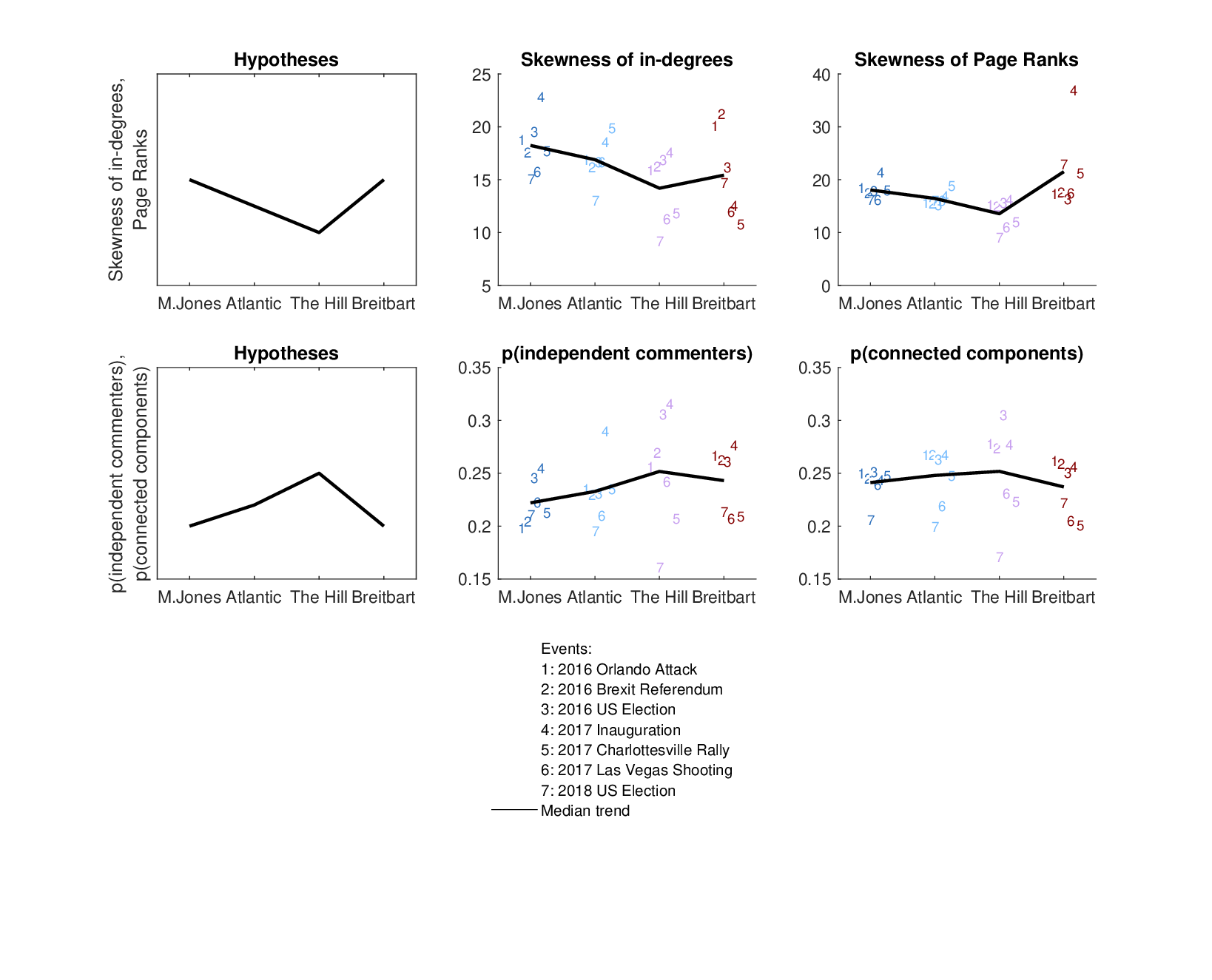}
        \caption{Differences in overall inequality of \deleted{commenters' influence}\added{attention} on different news websites, measured as the skew of centrality indices averaged across all commenters (compare with results for the top 1\% commenters in Figure~\ref{fig:inequal_1p_v2}). }
     \label{fig:inequal_all_v2}
\end{figure}

\begin{figure}[ht]
    \centering
        \includegraphics[width=1\textwidth]{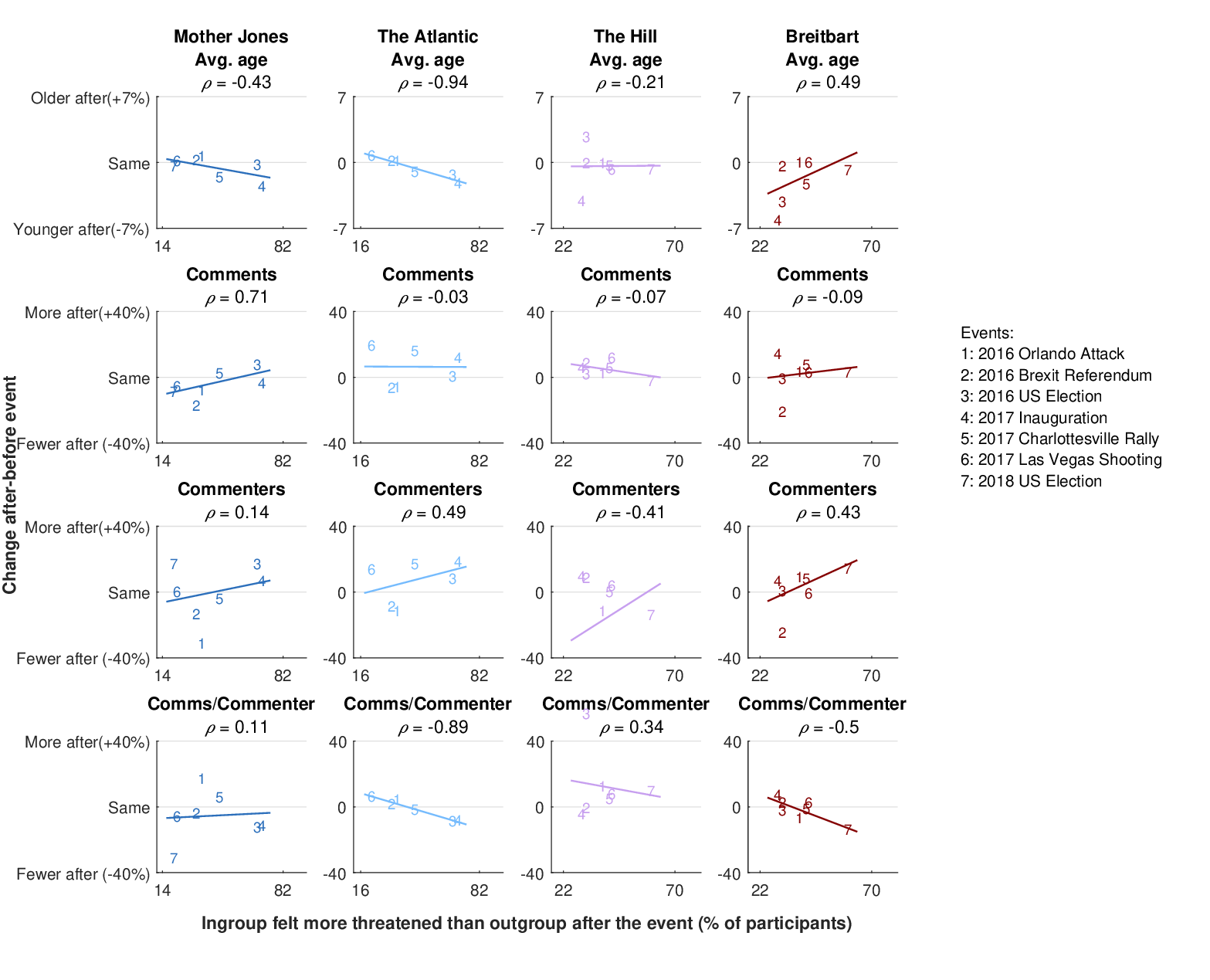}
        \caption{\added{Descriptive measures of comments and commentators for events rated as more or less threatening to the ingroups on different news sites. The threat ratings (x-axis) are summarized as the percentage of participants who estimated that the ingroup on a given site (Democrats for Mother Jones and The Atlantic, Republicans for The Hill and Breitbart) felt more threatened by the event than the outgroup (survey question 4, Figure~\ref{fig:threat_perception}). The first row shows the percent change in the average ``age" of commenters from before to after each event, where age is calculated as the difference in days between the event date and the date a commenter joined Disqus. The second row shows the percent change in the average number of comments per article from before to after each event. The third row shows the percent change in the average number of commenters per article from before to after each event. The fourth row shows the percent change in the ratio of comments to commenters from before to after each event. Colored numbers in each plot correspond to different events (see legend), and patterns are summarized by least-squares fitted lines across events and the Spearman $\rho$ value above each plot.}}
     \label{fig:desmes}
\end{figure}

\begin{figure}[ht]
    \centering
        \includegraphics[width=1\textwidth]{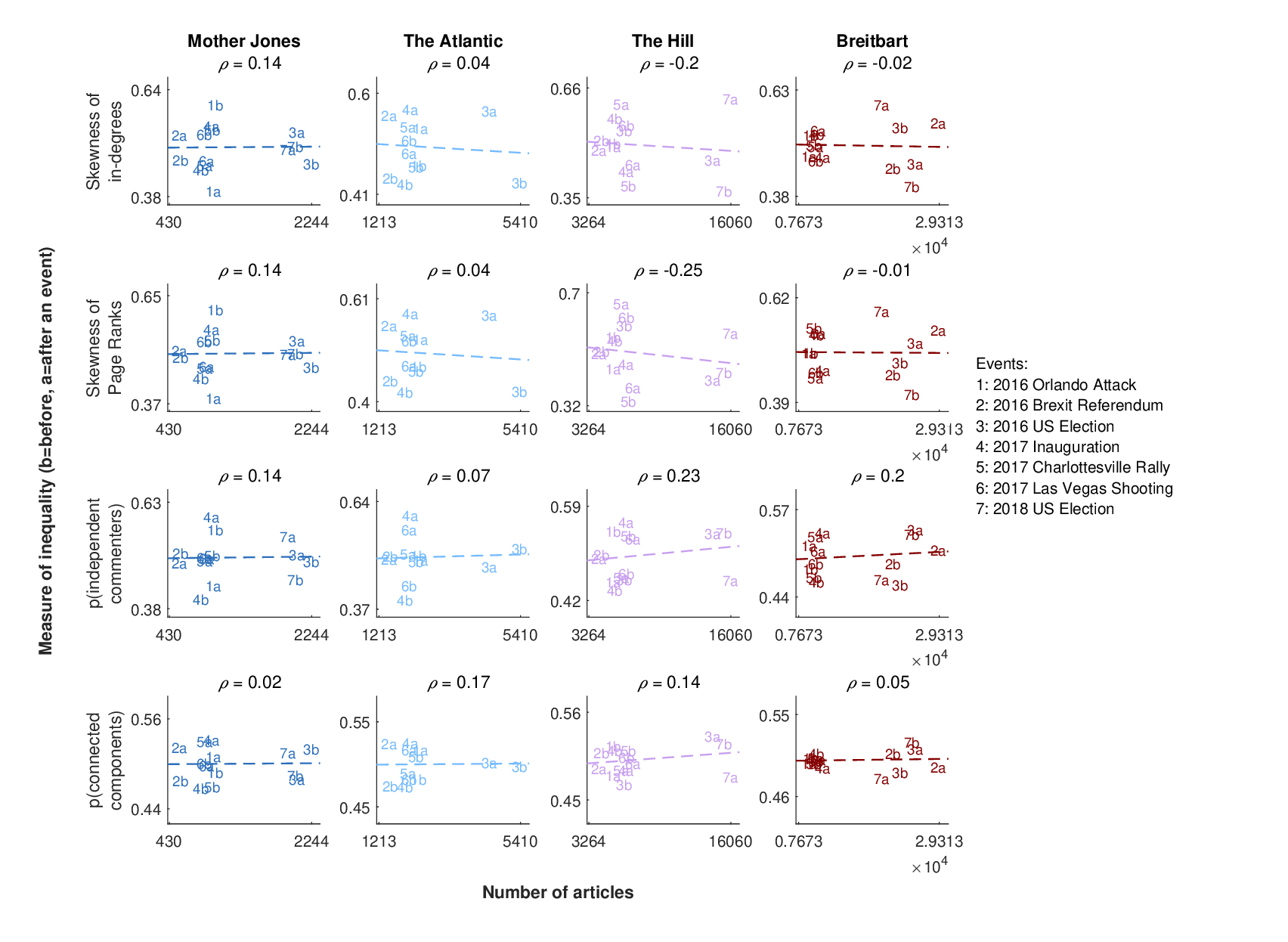}
        \caption{\added{The relationship of different measures of inequality (y-axes) and the number of articles published before and after each event (x-axes). Colored numbers in each plot correspond to different events (see legend), and patterns are summarized by least-squares fitted lines across events and the Spearman $\rho$ value above each plot.}}
     \label{fig:inequal_articles}
\end{figure}

\begin{figure}[ht]
    \centering
        \includegraphics[width=1\textwidth]{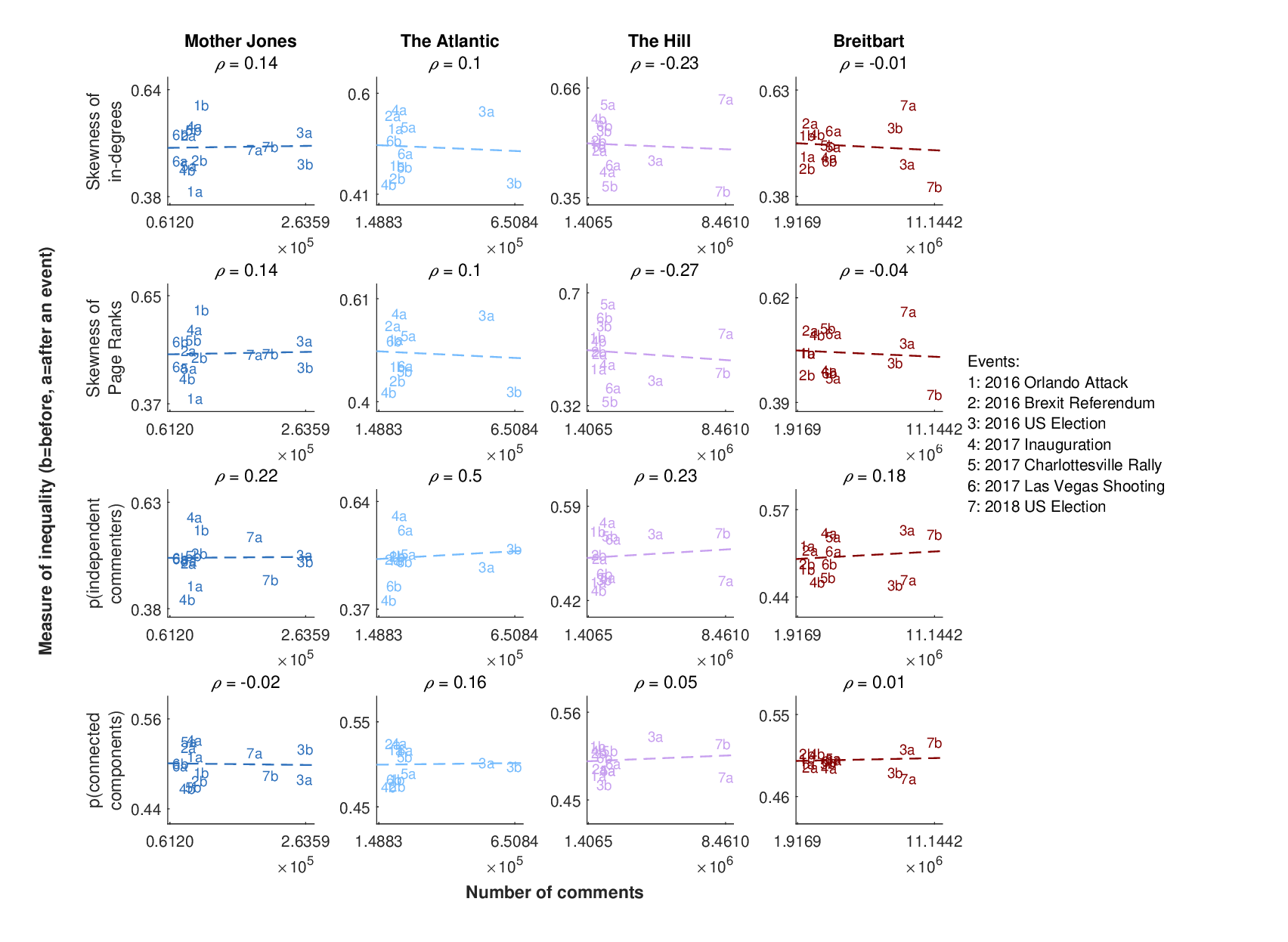}
        \caption{\added{The relationship of different measures of inequality (y-axes) and the number of comments posted before and after each event (x-axes). Colored numbers in each plot correspond to different events (see legend), and patterns are summarized by least-squares fitted lines across events and the Spearman $\rho$ value above each plot.}}
     \label{fig:inequal_comments}
\end{figure}

\begin{figure}[ht]
    \centering
        \includegraphics[width=1\textwidth]{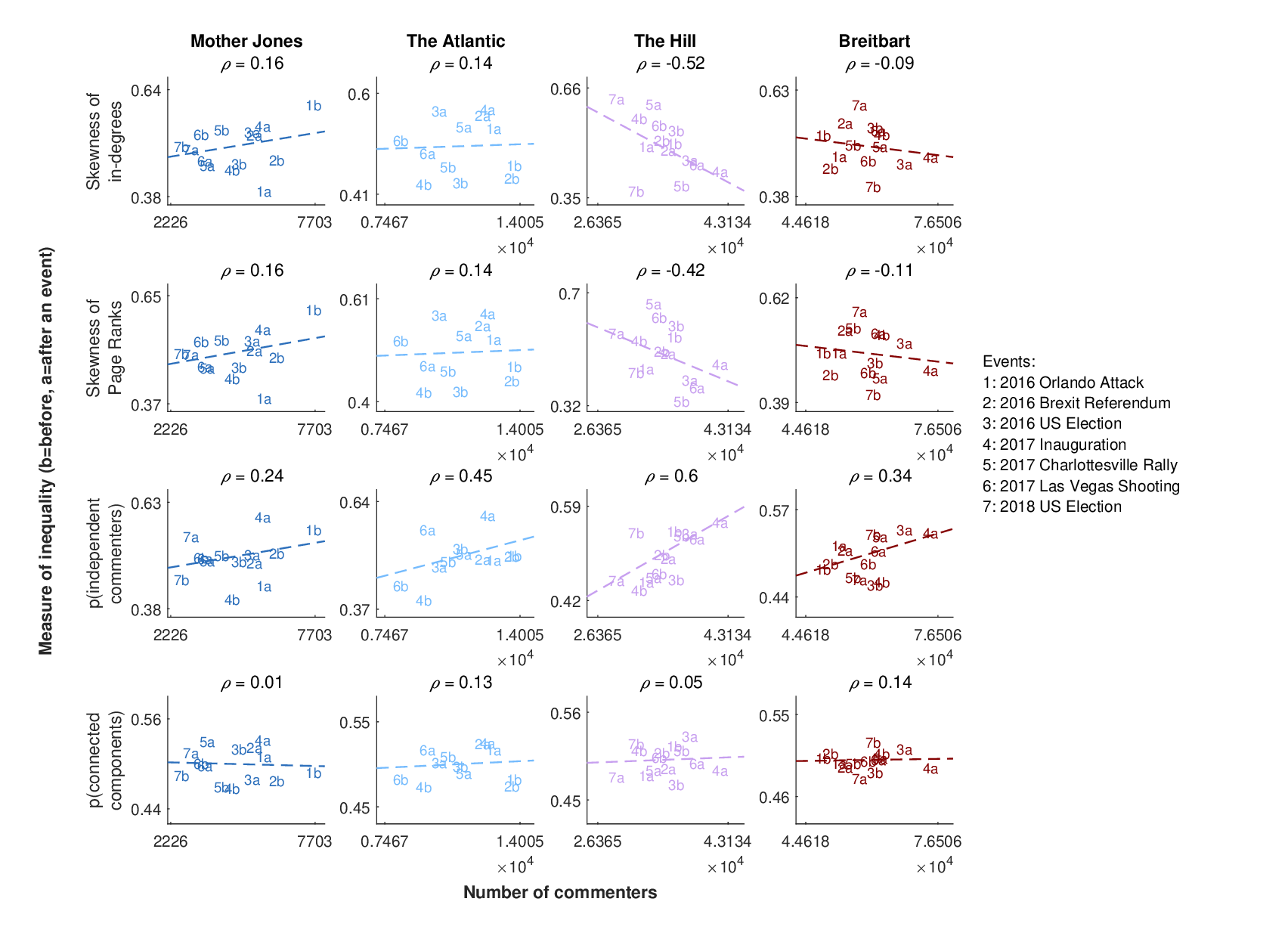}
        \caption{\added{The relationship of different measures of inequality (y-axes) and the number of commenters who participated in the discussions before and after each event (x-axes). Colored numbers in each plot correspond to different events (see legend), and patterns are summarized by least-squares fitted lines across events and the Spearman $\rho$ value above each plot.}}
     \label{fig:inequal_commenters}
\end{figure}

\begin{figure}[ht]
    \centering
        \includegraphics[width=1\textwidth]{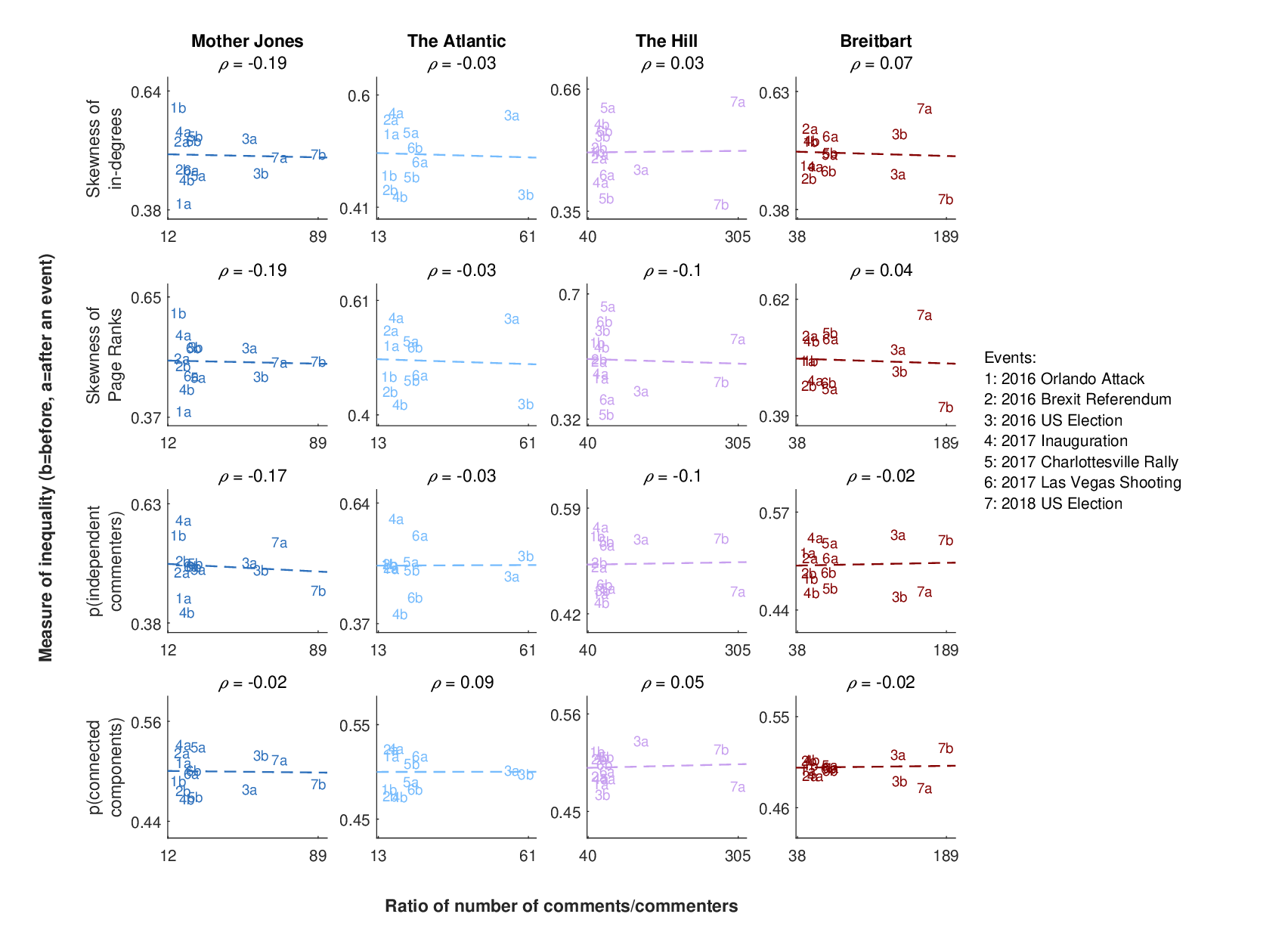}
        \caption{\added{The relationship of different measures of inequality (y-axes) and the ratio of comments to commenters before and after each event (x-axes). Colored numbers in each plot correspond to different events (see legend), and patterns are summarized by least-squares fitted lines across events and the Spearman $\rho$ value above each plot.}}
     \label{fig:inequal_comments_commenters}
\end{figure}


\end{document}